\documentclass[a4paper,fleqn,usenatbib,useAMS]{mnras}

\usepackage{graphicx}	
\usepackage{amsmath}	
\usepackage{amssymb}	
\usepackage{multicol}        
\usepackage{bm}		
\usepackage{pdflscape}	

\usepackage[flushleft]{threeparttable}

\newcommand{\gsim}{\;\lower.6ex\hbox{$\sim$}\kern-7.75pt\raise.65ex\hbox{$>$}\;}
\newcommand{\lsim}{\;\lower.6ex\hbox{$\sim$}\kern-7.75pt\raise.65ex\hbox{$<$}\;}

\newcommand {\kms} {km\,s$^{-1}$}
\newcommand {\Rmax} {R_\mathrm{max}}

\title[Globular clusters in NGC~4449]{LBT/MODS spectroscopy of globular clusters in the irregular galaxy NGC~4449.}

\author[F. Annibali et al.]{
F. Annibali,$^{1}$\thanks{E-mail: francesca.annibali@oabo.inaf.it}
E. Morandi,$^{2}$
L. L. Watkins,$^{3}$
M. Tosi,$^{1}$
A. Aloisi,$^{3}$
A. Buzzoni,$^{1}$ 
\newauthor
F. Cusano,$^{1}$  
M. Fumana,$^{4}$ 
A. Marchetti,$^{4}$ 
M. Mignoli,$^{1}$ 
A. Mucciarelli,$^{1,2}$ 
\newauthor
D. Romano,$^{1}$ 
and R. P. van der Marel,$^{3}$ 
\\
$^{1}$INAF- Osservatorio di Astrofisica e Scienza dello Spazio di Bologna, Via Piero Gobetti, 93/3, 40129 - Bologna, Italy\\
$^{2}$Dipartimento di Fisica e Astronomia, Universit\`a di Bologna, via Piero Gobetti 93/2, Bologna, Italy\\
$^{3}$Space Telescope Science Institute, 3700 San Martin Drive, Baltimore, MD 21218, USA\\
$^{4}$INAF-Istituto di Astrofisica Spaziale e Fisica Cosmica, Via Bassini 15, I-20133 Milano, Italy
}

\date{Accepted XXX. Received YYY; in original form ZZZ}

\pubyear{2018}

\begin{document}
\label{firstpage}
\pagerange{\pageref{firstpage}--\pageref{lastpage}}
\maketitle

\begin{abstract}

We present intermediate-resolution (R$\sim$1000) spectra in the $\sim$3\,500-10\,000 \AA\ range  of 14 globular  clusters in the magellanic irregular galaxy NGC~4449 acquired with the Multi Object Double Spectrograph on the Large Binocular Telescope. We derived Lick indices in the optical and the CaII-triplet index in the near-infrared in order to infer the clusters' stellar population properties. The inferred cluster ages are typically older than $\sim$9 Gyr, although ages are derived with large uncertainties. The clusters  exhibit 
intermediate metallicities, in the range $-1.2\lesssim$[Fe/H]$\lesssim-0.7$, and typically sub-solar [$\alpha/Fe$] ratios, with a peak at $\sim-0.4$. These properties suggest that i) 
during the first few Gyrs NGC~4449 formed stars slowly and inefficiently, with galactic winds having possibly contributed to the expulsion of the $\alpha$-elements, and ii) globular clusters in NGC~4449 formed relatively  ``late'', from a medium already enriched in the products of type Ia supernovae. The majority of clusters appear also under-abundant in CN compared to Milky Way halo globular clusters, perhaps because of the lack of a conspicuous N-enriched, second-generation of stars like that observed in Galactic globular clusters. 
Using the cluster velocities, we infer the dynamical mass of  NGC\,4449 inside 2.88~kpc to be M($<$2.88~kpc)=$3.15^{+3.16}_{-0.75} \times 10^9~M_\odot$. 
We also report the serendipitous discovery of a planetary nebula within one of the targeted clusters, a rather rare event.

\end{abstract}



\section{Introduction} \label{intro}

Dwarf galaxies are extremely important for our understanding of cosmology and  galaxy evolution for several reasons: 
i)  they are the most frequent type of galaxy in the Universe \citep[e.g.][]{lilly95}; 
ii) in the  Lambda Cold Dark Matter ($\Lambda$CDM) cosmological scenario, they are considered to be the first systems to collapse, supplying the building blocks for the formation of more massive galaxies through merging and accretion \citep[e.g.][]{kw93};
iii)  dwarf galaxies that experienced star formation (SF) at redshift z$>$6 are believed to be responsible for the re-ionization of the Universe \citep[e.g.][]{bouwens12};
iv) due to their low potential well, dwarf galaxies are highly susceptible to loss of gas and metals (either because of galactic outflows powered by supernova explosions or because of environmental processes) and thus provide an important contribution to the enrichment of the intergalactic medium (IGM).
Deriving how and when dwarf galaxies formed their stars is therefore fundamental to studies of galaxy formation and evolution. 

The most straightforward and powerful way to derive a galaxy's star formation history (SFH) is by resolving its stellar content: high signal-to-noise 
photometry of individual stars directly translates into deep color-magnitude diagrams (CMDs) that can be modeled to reconstruct the behavior of the star formation rate (SFR) as a function of look-back time \citep[e.g.][]{tosi91,bertelli92,gallart96,dolphin97,gro12,cignoni16}. Indeed, since the advent of the Hubble Space Telescope (HST), the study of resolved stellar populations in a large number of galaxies  within the Local Group and beyond has received a tremendous boost \citep[e.g.][]{annibali09,angst,mcquinn10a,annibali13,weisz14,sacchi16}. However, the look-back time that can be reached critically depends on the depth of the CMD: accurate information on the earliest epochs is only possible by reaching the oldest, faintest main sequence turnoffs and this translates, even with the superb spatial resolution of HST, into a typical distance of $\lesssim$1 Mpc from us. This distance limit corresponds in practice to a limit on the galaxy types for which the most ancient star formation history can be accurately recovered, the large majority of systems included in this volume being dwarf spheroidals (dSphs). 
With the obvious exception of the Magellanic Clouds, late-type, star-forming dwarfs are found at larger distances than dSphs, and the most active nearby star-forming dwarfs 
(e.g. NGC~1569, NGC~1705, NGC~4449) are as far as D$\gsim$3 Mpc.

Spectroscopic studies of (unresolved) star clusters provide an alternative approach to gather insights into the past SFH. Such studies are particularly valuable in those cases where the resolved CMD does not reach much below the tip of the red-giant branch (RGB), making it impossible to resolve the details of the SFH prior to 1$-$2 Gyr ago.  Indeed, star clusters are present in all types of galaxies and are suggested to be the birth site of many (possibly most) stars. The formation of massive star clusters is thought to be favored by the occurrence of intense star formation events, as suggested by the presence of a correlation between the cluster formation efficiency and the galaxy SFR density \citep[e.g.][]{larsen00,goddard10,adamo11b,adamo15}. Therefore, clusters can be powerful tracers of the star formation process in their host galaxies.

Despite the large number of photometric studies of clusters in dwarf galaxies performed so far \citep[e.g.][]{hunter00,hunter01,billett02,annibali09,adamo11a,anni11,cook12,pellerin12}, only a few systems (early-type dwarfs, in the majority of cases) have been targeted with 6-10 m class telescopes to derive cluster spectroscopic ages and chemical composition 
\citep[e.g.][]{puzia00,strader03,strader05,conselice06,sharina07,sharina10,strader12}. Although spectroscopic ages older than $\sim$2 Gyr are derived with large uncertainties, 
both because of the well known age-metallicity degeneracy and because of the progressively lower age-sensitivity of the Balmer absorption lines with increasing look-back time,
the additional information on the chemical abundance ratios can provide stringent constraints on the galaxy SFH back to the earliest epochs. 

In this paper we present deep spectroscopy obtained with the Large Binocular Telescope (LBT) of clusters in the Magellanic irregular galaxy NGC~4449 ($\alpha_{2000}$=$12^h 28^m 11^{s}.9$ $\delta_{2000}$=$+44^{\circ} 05^{'} 40^{"}$) at a distance of $3.82 \pm 0.18$ Mpc from us \citep{annibali08}. With an integrated 
absolute magnitude of $M_B$=$-18.2$, NGC~4449 is $\approx$1.4 times more luminous than the Large Magellanic Cloud (LMC). Its metallicity has been derived through spectroscopy of H~II regions and planetary nebulae \citep[e.g.][] {berg12,pily15,annibali17}, and ranges between  12 + $\log$(O/H) = 8.26 $\pm$ 0.09 and  12 + $\log$(O/H) = 8.37 $\pm$ 0.05, close to the LMC value, although \citet{kumari17}  found a metallicity as low as 12 + $\log$(O/H) = 7.88 $\pm$ 0.14 in the very central galaxy regions.  
NGC~4449 is remarkable for several reasons: it is  one of 
the most luminous and active nearby irregular galaxies,  with a current SFR of $\sim 1$ M$_{\odot}$ yr$^{-1}$  \citep{mcquinn10b,sacchi17}; it has a conspicuous population of clusters ($\approx$80), with a specific frequency of massive clusters higher than in nearby spirals and in the LMC \citep{anni11}; it hosts an old, very massive and elliptical cluster, associated with two tails of young stars, that has been suggested to be the nucleus of a former gas-rich satellite galaxy undergoing tidal 
disruption by NGC~4449 \citep{annibali12}; it is the first dwarf galaxy where a stellar tidal stream has been discovered \citep[][]{delgado12,rich12}; 
it has a very extended HI halo ($\sim 90$~kpc in diameter) which is a factor of $\sim 10$ 
larger than the optical diameter of the galaxy, and that appears to rotate in the opposite direction to the gas in the center  \citep{hunter98}.  All these studies suggest that NGC~4449 experienced a complex evolution and was possibly built, at least in part, through the accretion of satellite galaxies.

From optical CMDs of the stars resolved with the advanced Camera for Survey (ACS) on board HST, \citet{mcquinn10b} and \citet{sacchi17} derived the SFH of NGC~4449. These analyses indicate that NGC~4449 enhanced its star formation activity $\sim$500 Myr ago, while the rate was much lower at earlier epochs; however, the impossibility of reaching the old main-sequence turnoffs or even the red clump/horizontal branch  with the available data implies that the SFH of NGC~4449 is very uncertain prior to $\sim$1-2 Gyr ago. Star clusters appear mostly unresolved in NGC~4449; \citet{anni11} performed integrated-light photometry of $\sim$80 young and old clusters identified in the  ACS images, and found that their colors are compatible, under the assumption of a metallicity of $\sim$1/4 solar, with a continuous age distribution over the whole  Hubble time. However, only spectroscopy can allow for real progress by breaking, to some extent,   the age-metallicity degeneracy and by providing information on the element abundance ratios. Motivated by this goal, we performed a spectroscopic follow-up of a few clusters in the \citet{anni11} sample using the Multi Object Double Spectrographs on the Large Binocular Telescope (LBT/MODS). In Section~\ref{data_reduction} we describe the observations and the data reduction; in Section~\ref{index_section} we compute optical and near-infrared absorption-line indices;  in Section~\ref{stpop} we derive the cluster stellar population parameters; in Section~\ref{cl58_section} we present our serendipitous discovery of a candidate PN within one of the clusters; in Section~\ref{dynamics} we obtain an estimate of the dynamical mass of NGC~4449 from cluster velocities; in Sections~\ref{discussion} and ~\ref{conclusions} we discuss and summarize our results.

\begin{figure*}
	\includegraphics[width=\textwidth]{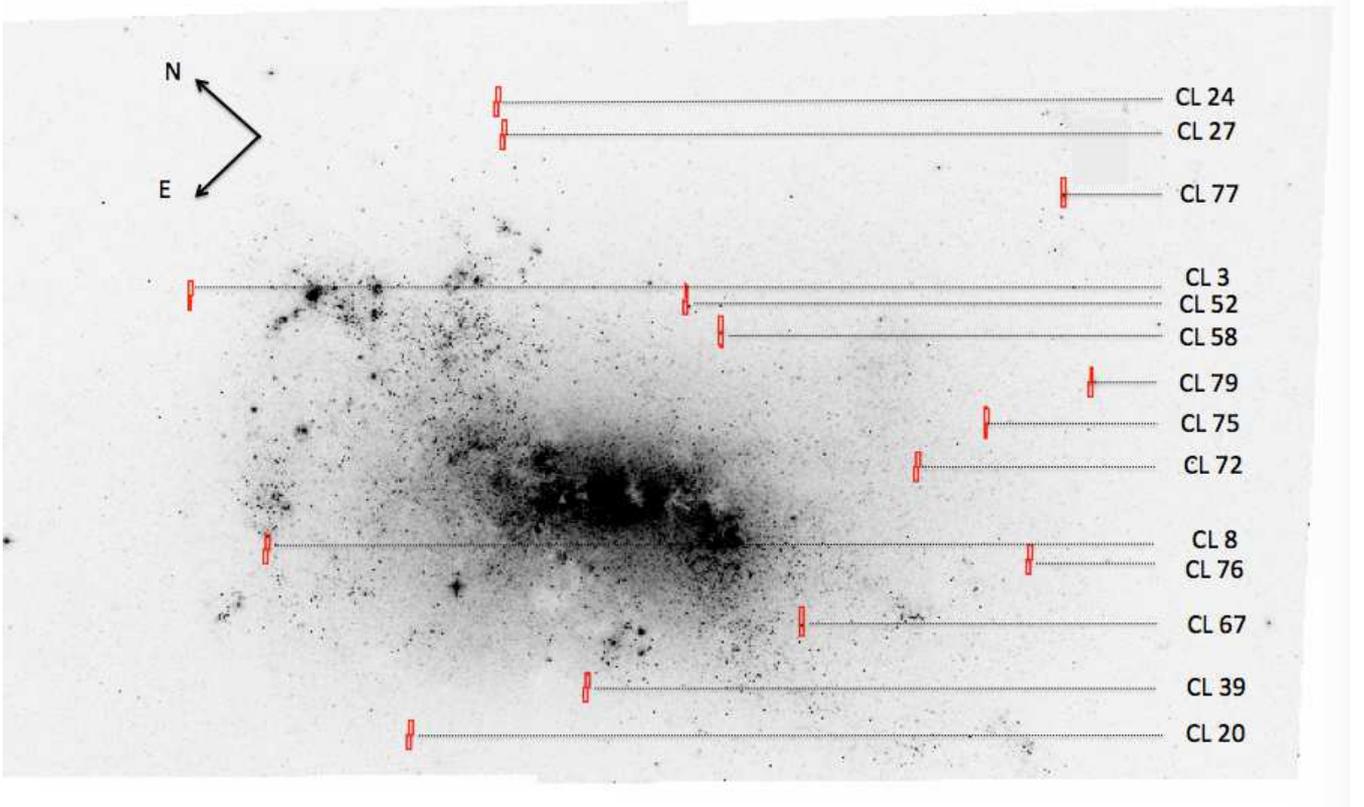}
    \caption{HST/ACS mosaic image of NGC~4449 in F555W ($\sim$V) with superimposed the LBT/MODS 1''$\times$8''  slits centered on star clusters.  The 
cluster identification numbers correspond to those in \citet{anni11}.}
    \label{image}
\end{figure*}

\section{Observations and data reduction} \label{data_reduction}

\begin{figure*}
	\includegraphics[width=\textwidth]{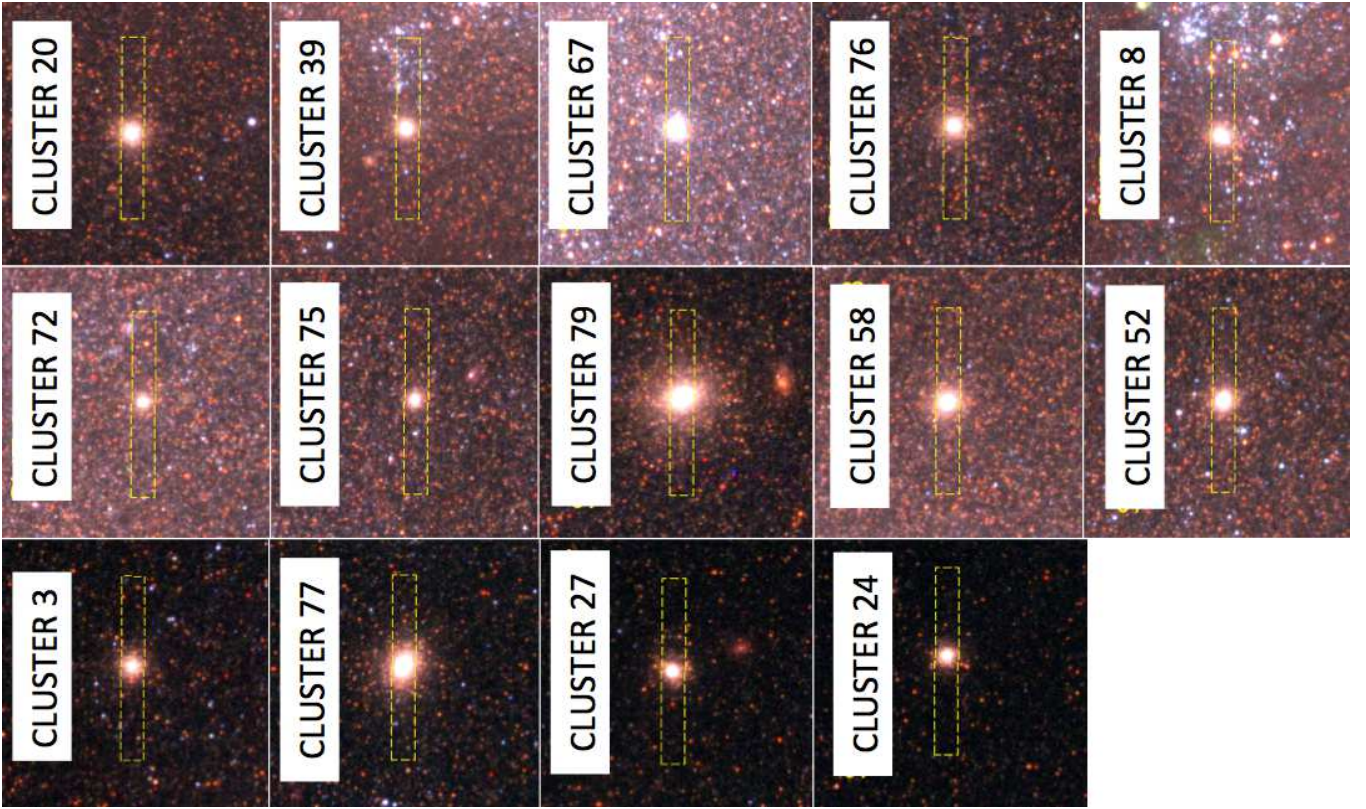}
    \caption{HST/ACS color-combined images  (F435W=blue, F555W=green, F814W=red)  of clusters in NGC~4449 targeted for spectroscopy 
with LBT/MODS. In each box, the cluster identification number  is that of  \citet{anni11}. Each field of view is $12'' \times 12''$, and the slit is 1''$\times$8''.}
    \label{mosaic}
\end{figure*}

Clusters to be targeted for spectroscopy were selected from the  catalog of  \citet{anni11}, hereafter A11, based on the HST/ACS F435W (B), F555W (V), and F814W (I) images acquired within GO program 10585, PI Aloisi.  
We selected 14  clusters located in relatively external regions of NGC~4449 to avoid the severe crowding toward the most central star-forming regions. 
The observations were performed with  the Multi Object Double Spectrographs on the Large Binocular Telescope (LBT/MODS) in January 21 and 22, 2013 within program 2012B\_23, run~B (PI Annibali). The 1''$\times$8'' slit mask superimposed on the ACS V image is shown in Figure~\ref{image}, while Figure~\ref{mosaic} shows color-composite ACS images for the target clusters.  
The observations were obtained with the blue G400L (3200$-$5800 \AA) and the red G670L (5000$-$10000 \AA) gratings on the blue and red channels in dichroic mode 
for  8$\times$1800 sec, for a total integration time of 4 h. Notice that the instrumental sensitivity in dichroic mode  is very low in the $\sim$5500-5800 \AA \ range.
The seeing varied from $\sim$0.7'' to $\sim$0.9'', and the average airmass  was $\sim$1.4. 
The journal of the observations is provided in Table~\ref{obs}.  
Three Lick standard stars of F-K spectral types (see Table~\ref{lick_std}) were also observed during our run with a  1''$\times$8'' longslit with the purpose of calibrating our measurements into the widely used Lick-IDS system (see Section \ref{index_section} for more details).

\begin{table}
	\caption{Journal of LBT/MODS observations for clusters in NGC~4449}
	\label{obs}
	\begin{threeparttable}
	\begin{tabular}{lcccccc} 
\hline
N. & Date-obs  & Exptime   & Seeing   & Airmass & PosA & ParA   \\
\hline
1 &  2013-01-20   & 1800 s  & 0.9''  & 1.7   & $-35^{\circ}$ & $-78^{\circ}$ \\
2 &  2013-01-20   & 1800  s & 0.9''  & 1.5   & $-35^{\circ}$ & $-83^{\circ}$ \\
3 &  2013-01-20   & 1800  s & 0.8''  & 1.3   & $-35^{\circ}$ & $-88^{\circ}$ \\
4 &  2013-01-20   & 1800 s & 0.7''   & 1.2   & $-35^{\circ}$ & $-93^{\circ}$ \\
5 &  2013-01-21   & 1800 s & 0.9''   & 1.7   & $-35^{\circ}$ & $-76^{\circ}$ \\
6 &  2013-01-21   & 1800 s & 0.8''   & 1.5   & $-35^{\circ}$ & $-81^{\circ}$ \\
7 &  2013-01-21   & 1800 s & 0.8''   & 1.4   & $-35^{\circ}$ & $-86^{\circ}$ \\
8 &  2013-01-21   & 1800 s & 0.7''   & 1.3   &  $-35^{\circ}$ & $-91^{\circ}$ \\
\hline
\end{tabular}
\begin{tablenotes}
\small
      \item Col.~(1): exposure number; Col. (2): date of observations; Col. (3): exposure time in seconds; Col. (4): average seeing in arcsec; 
      Col (5): average airmass; Col (6): position angle in degree; Col (7): parallactic angle in degree. 
    \end{tablenotes}
\end{threeparttable}
 \end{table}

\begin{table}
	\centering
	\caption{Observed Lick standard stars.}
	\label{lick_std}
	\begin{threeparttable}
		\begin{tabular}{lcccc} 
\hline
Name & Spectral Type  & Date-obs   & Exptime   & V \\
\hline
HD~74377 & K3~V & 2013-01-20 & 1 s & 8.2 \\
HD~84937 & F5~VI & 2013-01-20 & 1 s & 8.3 \\
HD~108177 & F5~VI & 2013-01-21 & 1 s & 9.7 \\
\hline
\end{tabular}
\begin{tablenotes}
\small
\item Lick standard stars were observed with a 1''$\times$8'' longslit; Col.~(1): star name; Col. (2): spectral type; Col. (3): date of observations; Col. (4): exposure time in seconds; Col. (5): V apparent magnitude.
    \end{tablenotes}
\end{threeparttable}
\end{table}

Bias and flat-field subtraction, and wavelength calibrations were performed  with the Italian LBT Spectroscopic reduction Facility at INAF-IASF Milano, 
producing the calibrated two-dimensional (2D) spectra for the individual sub-exposures. 
The accuracy of the wavelength calibration obtained from the pipeline was checked against prominent sky lines adopting the nominal sky wavelengths tabulated in \citet{uves_sky}. In fact, while arc-lamps typically provide a good calibration of the wavelength variation with pixel position, a zero-point offset may be expected due to the fact that the light from the night sky and the light from the lamp do not follow the same path through the optics. In our case, we found that the $\Delta \lambda$ offset depended on slit position and was lower than $\sim$1 \AA \ for both the blue and the red spectra;  this zero-point correction was applied  to our data. 

\begin{figure}
	\includegraphics[width=\columnwidth]{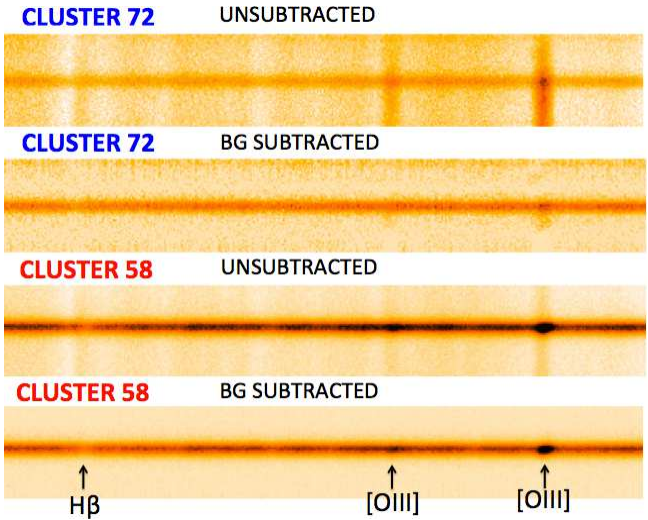}
    \caption{Two examples of cluster spectra contaminated by emission lines. Portions of the spectra around the H$\beta$ and [O~III]$\lambda\lambda$4959,5007 \AA\ lines are shown. For each cluster, the total-combined un-subtracted  spectrum is in the higher row, while the background-subtracted spectrum is just below. Cluster CL~72 suffers contamination from the diffuse ionized  gas in NGC~4449, and some residual emission is still present after background subtraction. In cluster CL~58, the 
(modest) contamination from the diffuse ionized gas in NGC~4449 is optimally removed with the background subtraction; on the other hand, the cluster exhibits a quite strong 
centrally concentrated emission.}
    \label{bg_sub}
\end{figure}

Sky subtraction was performed on the 2D calibrated spectra; to this purpose, we used the {\it{background}} task in IRAF\footnote{ IRAF is distributed by the National Optical Astronomy Observatory, which is operated by the Association of Universities for Research in Astronomy, Inc., under cooperative agreement with the National Science Foundation.}, typically choosing the windows at the two opposite sides of the cluster.  This procedure does in principle remove, together with the sky, also the contribution from the NGC~4449' s unresolved background. In particular,  the {\it{background}} subtraction  should remove from the cluster spectra possible emission line contamination due to the presence of 
diffuse ionized gas in NGC~4449;  however, as we will see in Section~\ref{em_corr}, the subtraction is not always perfect (e.g. because of the highly variable emission background) and some residual emission may still be present in some clusters after background subtraction. This is shown in Figure~\ref{bg_sub}, where we present the case of cluster CL~72  as an illustrative example. Here, the emission line spectrum is highly variable within the slit, preventing a perfect background removal; as a result, the final background-subtracted spectrum appears still contaminated by residual emission lines.  

Ionized gas emission is detected in all the clusters of our sample except clusters CL~75, CL~79, and CL~ 77; as expected, the strongest emission is observed for the clusters located in the vicinity of star forming regions, i.e. clusters CL~39, CL~67, and CL~8. Cluster CL~58 shows strong emission as well but, at variance with all the other clusters, its emission appears ``nuclear'', i.e. confined within the cluster itself (see Fig.~\ref{bg_sub}). We will come back to the case of cluster CL~58 later in Section  \ref{cl58_section}.

We combined the sky-subtracted 2D spectra into a single frame for the blue and the red channels, respectively. 
The  {\it apall} task in the {\it twodspec} IRAF package was then used to extract the 1D spectra from the 2D ones. In order to refine the emission subtraction, the {\it apall} task was run with the background subtraction option ``on''. To derive the effective spectral resolution, we used the combined 1D spectra with no sky subtraction, and measured the FWHM of the most prominent sky lines; this resulted into resolutions of R$\sim$1040 and R$\sim$1500  at 4358 \AA \ and 7000 \AA, respectively.

The blue and red 1D spectra were flux calibrated using the sensitivity curves from the Italian LBT spectroscopic reduction pipeline; the curves  were derived using the spectrophotometric standard star Feige~66 observed  in dichroic mode with a 5''-width  slit  on January 20, 2013.  To obtain the red and blue sensitivity curves, the observed standard was compared with reference spectra in the HST CALSPEC database.  Atmospheric extinction corrections were applied using the average extinction curve available from the MODS calibration webpage at http://www.astronomy.ohio-state.edu/MODS/Calib/. As discussed in our study of H~II regions and PNe in NGC~4449 based on LBT/MODS data acquired within run~A of the same program \citep{annibali17}, this may introduce an uncertainty in flux calibration as  high as $\sim15\%$ below $\sim$ 4000 \AA. 

\subsection{Differential Atmospheric Refraction} \label{daf_section}

\begin{figure*}
	\includegraphics[width=\textwidth]{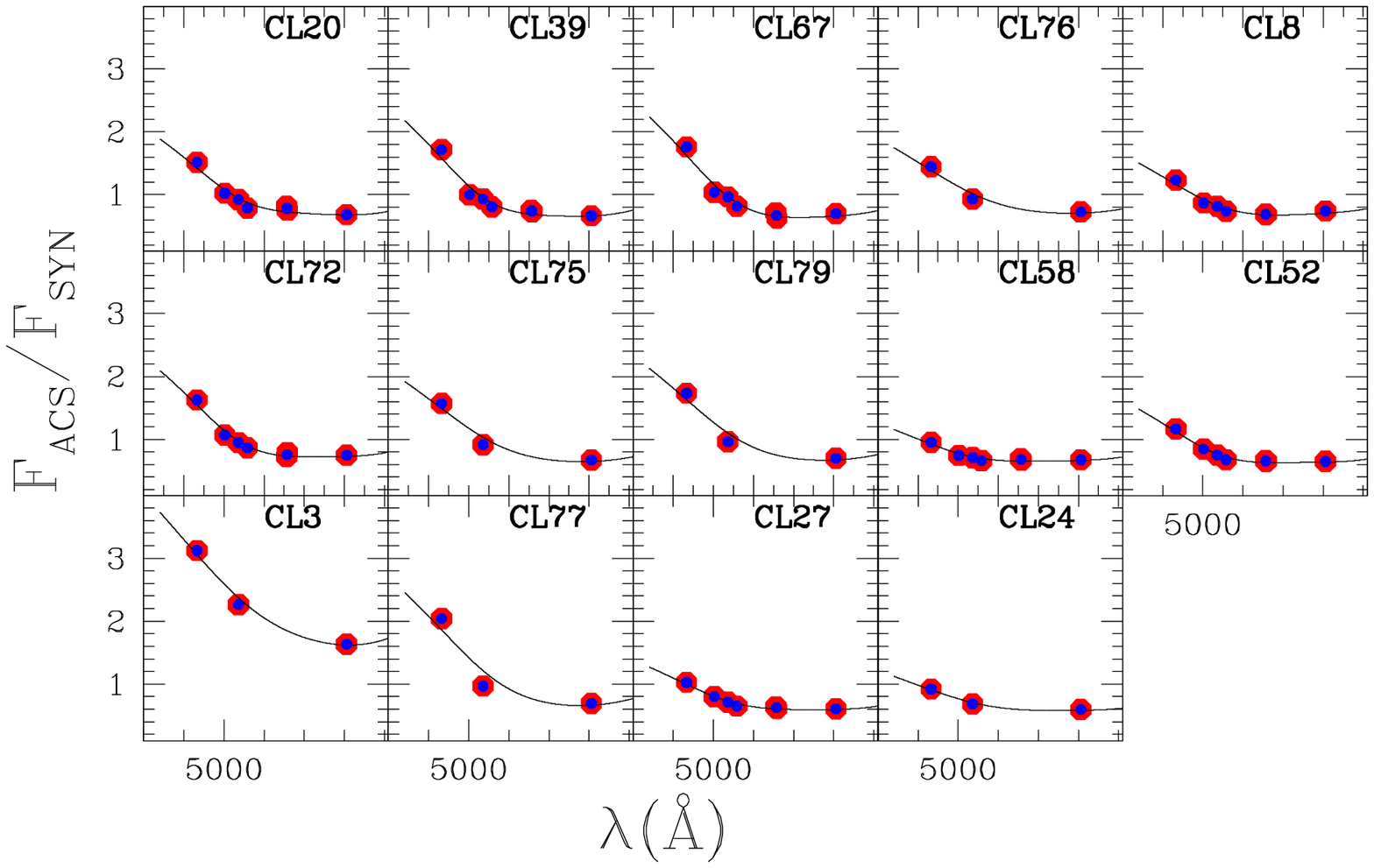}
\caption{Effect of differential atmospheric refraction on our cluster spectroscopy. In ordinate, $F_{ACS}/F_{SYN}$ is the ratio between the measured ACS fluxes (in F435W, F555W, F814W, F502N, F550M, F658N, and F660N) and the  Synphot  synthetic  fluxes  (see Section~\ref{data_reduction} for details). 
The curve is the fit obtained adopting a radial basis function (RBF).
\label{calibration}}
\end{figure*}

Our observations were significantly affected by differential atmospheric refraction \citep{filippenko82} due to the displacement between the slit position angle and the parallactic angle, coupled with the relatively high airmass (see the journal of observations in Table~\ref{obs}). In order to quantify this effect as a function of wavelength, we used HST/ACS imaging in F435W (B), F555W (V), F814W (I), and F658N (H$\alpha$) from our GO program 10585  and archive HST/ACS imaging in F502N (O~III), F550M and F660N (H$\alpha$) 
from GO program 10522 (PI Calzetti). The F435W, F555W and F814W images cover a field of view of $\sim380''\times200''$, obtained with two ACS pointings, while the  F502N, F550M, F660N and F658N images have  a smaller field of view of $\sim$200''$\times$200''. As a consequence, the most central clusters have photometric coverage in all seven bands, while more ``peripheral'' ones (i.e. clusters CL~76, CL~75, CL~79, CL~3, CL~77, and CL~74) are detected only in B, V and I. 
Aperture photometry of the clusters was performed on the ACS images  with the {\it Polyphot} IRAF task  adopting a polygonal aperture resembling the same aperture used to extract the 1D spectra from the 2D ones. Synthetic magnitudes in the same ACS filters were then derived  with the {\it Calcphot} task in  the {\it Synphot} package by convolving the MODS spectra with the ACS bandpasses. Figure~\ref{calibration} shows the ratio of the ACS fluxes over the synthetic {\it Synphot} fluxes as a function of wavelengths. The fit to the data points was obtained adopting a radial basis function (RBF) available in Python 
\footnote{RBF=$\sqrt{(r/\epsilon)^2 + 1}$, where $r$ is the distance between any point and the centre  of the basis function; 
see http://docs.scipy.org/doc/scipy/reference/generated/scipy.interpolate. Rbf.html.}. 
Fig.~\ref{calibration} shows a significant increase of the $F_{ACS}/F_{SYN}$ ratio toward bluer wavelengths, while the trend flattens out in the red. This behaviour translates into a
loss of flux in the blue region of our spectra, as expected in the case of a significant effect from differential atmospheric refraction. We also notice that, at red wavelengths, the $F_{ACS}/F_{SYN}$ ratio is not equal to unity but lower than that, indicating that the fluxes from the spectra are overestimated. As discussed in \citet{smith07} 
and \citet{annibali17}, flux calibrations are usually tied to reference point source observations, and therefore include an implicit correction for the fraction of the point source light that falls outside the slit; however, in the limit of a perfect uniform, slit-filling extended source, the diffractive losses out of the slit are perfectly balanced by the diffractive gains into the slit from emission beyond its geometric boundary; therefore, point-source-based calibrations likely cause an overestimate of the extended source flux. We used the curves obtained from  the fits in Fig.~\ref{calibration} to correct our spectra; however, it is worth noticing that the line strength indices that will be used in Section~\ref{index_section} do not depend on absolute flux calibration (since they are normalized to local continua) and depend only very slightly on relative flux calibration as a function of wavelength.

\subsection{Radial Velocities}

\begin{figure}
\includegraphics[width=\columnwidth]{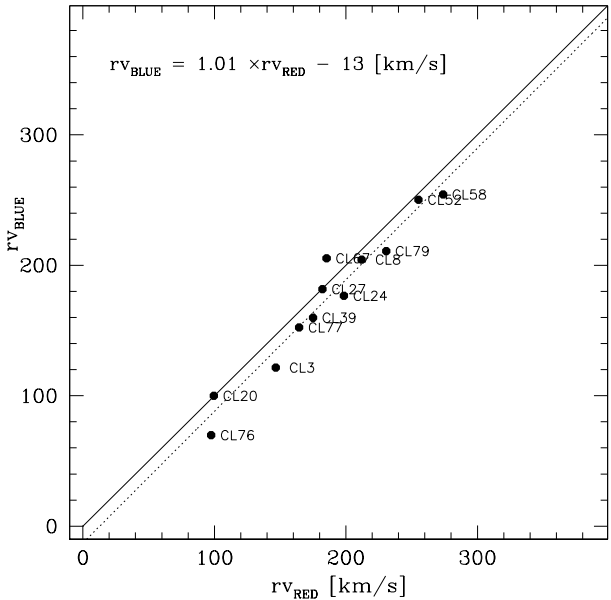}
\caption{Comparison between cluster radial velocities obtained with the blue and with the red MODS spectra, respectively. The solid line is the one-to-one relation, while the 
dotted line is the linear fit to the data points. Notice that the displayed velocities have not been corrected here for the motion of the Sun.  
\label{vel_cfr}}
\end{figure}

Radial velocities for the clusters were derived through the {\it fxcor} IRAF task by convolving the cluster spectra both with theoretical simple stellar populations and with observed stellar spectra.  In the blue, the convolution was performed within the $\sim$3700-5500 \AA \ range, masking, when present, emission lines; in the red, we used the CaII triplet in the range $\sim$8000-8400 \AA. The velocities measured from the blue and red spectra present a systematic offset of $\lesssim$10 km/s, as shown in Figure~\ref{vel_cfr}. 
This sets an upper limit to spurious sources of error, such as the differential atmospheric refraction already discussed in Section~\ref{daf_section}: indeed, the fact that the clusters are mis-centered in the slit as a function of wavelength causes also a small velocity shift. However, Figure~\ref{vel_cfr} demonstrates that this error is modest and that the derived velocities are suitable for a dynamical analysis (see Section~\ref{dynamics}). We find that the velocity uncertainties provided by the  {\it fxcor} task are lower for the red than for the blue spectra, likely because of the higher signal-to-noise of the CaII triplet lines compared to the lines used in the blue spectral region. Notice also that in general the CaII triplet lines are easier to use for velocity measurements than blue lines, e.g. because they are less sensitive to template mismatch. Furthermore, they are better calibrated in wavelengths thanks to the wealth of sky lines in the red spectral region. For all these reasons, we elect hereafter the cluster velocities obtained from the CaII triplet for our study.

The velocities are provided in Table~\ref{rvel}.  Corrections for the motion of the Sun were performed with the  {\it rvcorrect} task in IRAF. In addition, we provide radial velocities for the NGC~4449's H~II regions and PNe observed during run~A.  The H II region and PN velocities were derived from the most prominent emission lines, and also in this case the absolute wavelength calibration was anchored to strong sky lines observed in the spectra. The cluster radial velocities will be used in Section~\ref{dynamics} to infer basic dynamical properties in NGC~4449.

\begin{table}
	\centering
	\caption{Heliocentric radial velocities of clusters, H~II regions and PNe in NGC~4449.}
	\label{rvel}
	\begin{threeparttable}
	\begin{tabular}{lccc} 
\hline
Name & R.A. (J2000) & Dec (J2000)  & Vhelio \\
 & [hh mm ss] & [dd mm ss]  & [km/s] \\
\hline
 CL~79   &   12  27  59.87  & 44  04  43.34   &  246$\pm$15   \\
 CL~58   &  12  28  06.44   & 44  05  55.12   &  289$\pm$14   \\
 CL~8    &   12  28  18.79   & 44  06  23.08   & 227$\pm$20  \\
 CL76     &   12  28  03.88  & 44  04  15.26   & 113$\pm$ 20  \\
 CL~67   &  12  28  09.34   & 44  04  38.99   &  201$\pm$22  \\
 CL~39   &  12  28  14.60   & 44  05  00.43   &  190$\pm$15   \\
 CL~20   &  12  28  18.82   & 44  05  19.61   & 115$\pm$15   \\
 CL~52  &  12  28  06.64    & 44  06  07.78   &  270$\pm$ 14   \\
 CL~3    &   12  28  16.44   & 44  07  29.49   & 162$\pm$23   \\
 CL~77  &   12  27  57.53   & 44  05  28.16   &  180$\pm$14   \\
 CL~27  &  12  28  07.72   & 44  07  13.42    & 197$\pm$23   \\
 CL~24  &   12  28  07.32  & 44  07  21.54    & 214$\pm$26   \\
 \hline
 PN~1    &  12 28 04.126 &  44 04 25.14 & 225$\pm$14 \\
PN~2     &  12 28 03.540 &  44 04 34.80 & 196 $\pm$10 \\
PN~3     &  12 28 03.972 &  44 05 56.78 &  173 $\pm$9  \\
H~II-1    &  12 28 12.626 &  44 05 04.35  & 191$\pm$ 7 \\
H~II-2    &  12 28 09.456 &  44 05 20.35 & 168$\pm$5  \\
H~II-3    &  12 28 17.798 &  44 06 32.49 & 225$\pm$8  \\
H~II-4    &  12 28 16.224  & 44 06 43.32 & 218$\pm$23 \\ 
H~II-5    &  12 28 13.002 &  44 06 56.38 & 217$\pm$16  \\
H~II-6    &  12 28 13.925 &  44 07 19.04  & 232$\pm$ 4   \\
\hline
\end{tabular}
\begin{tablenotes}
\small
\item 
Col.~(1): Cluster, H~II region or PN identification; Col. (2-3): right ascension and declination in J2000; Col. (4): heliocentric velocities in km/s.
\end{tablenotes}
\end{threeparttable}
\end{table}

\subsection{Final Spectra}
   
\begin{figure*}
\includegraphics[width=\textwidth]{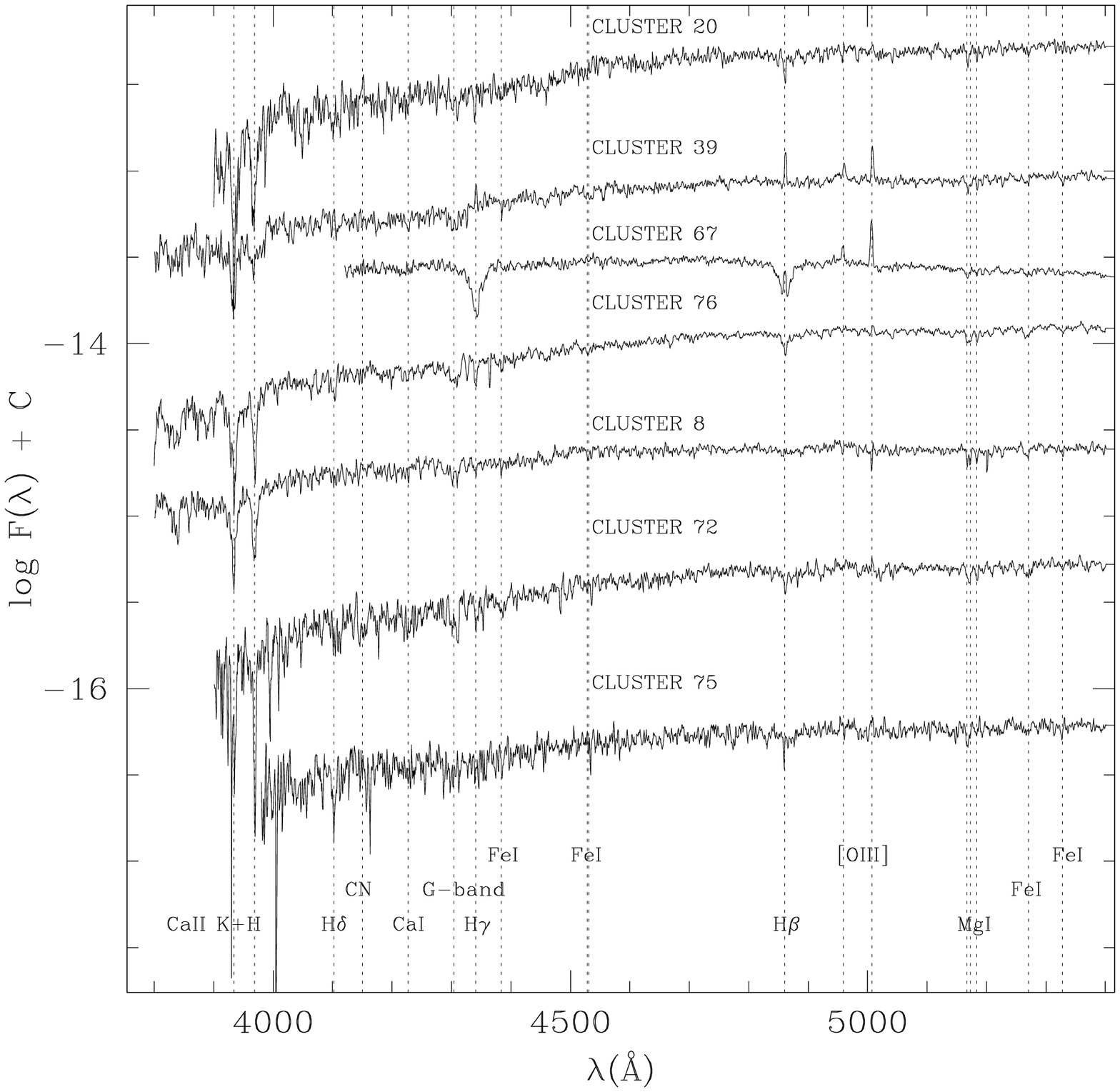}
\caption{LBT/ MODS blue calibrated spectra in the rest-frame for clusters CL~20, CL~39, CL~67, CL~76, CL~8, CL~72 and CL~75 in NGC~4449. 
Indicated are the most prominent absorption and emission lines. 
\label{spectra_blue_a}}
\end{figure*}

\begin{figure*}
\includegraphics[width=\textwidth]{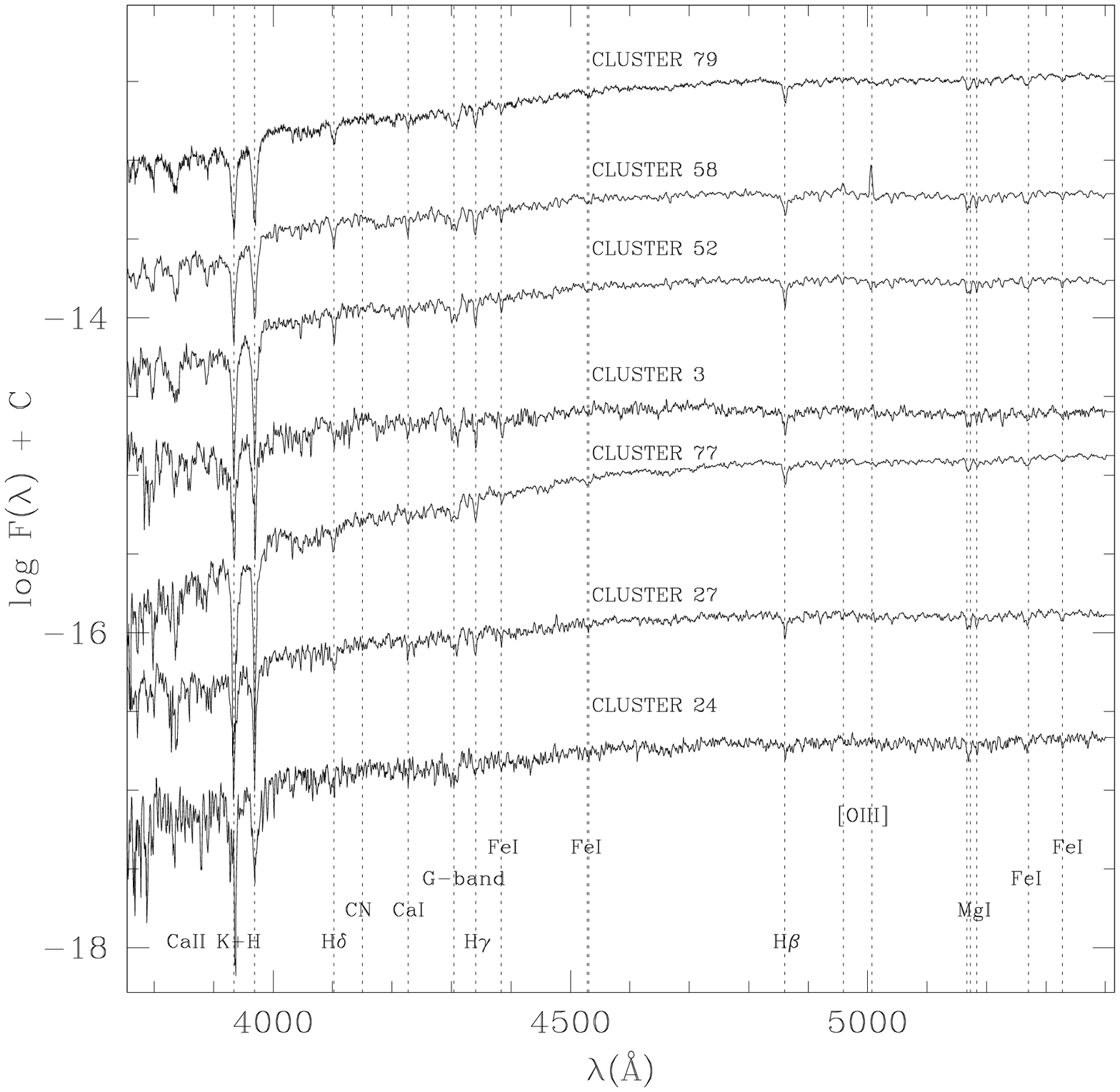}
\caption{Same as in Fig.~\ref{spectra_blue_a} but for clusters CL~79, CL~58, CL~52, CL~3, CL~77, CL~ 27 and CL~24. 
\label{spectra_blue_b}}
\end{figure*}

\begin{figure*}
\includegraphics[width=\textwidth]{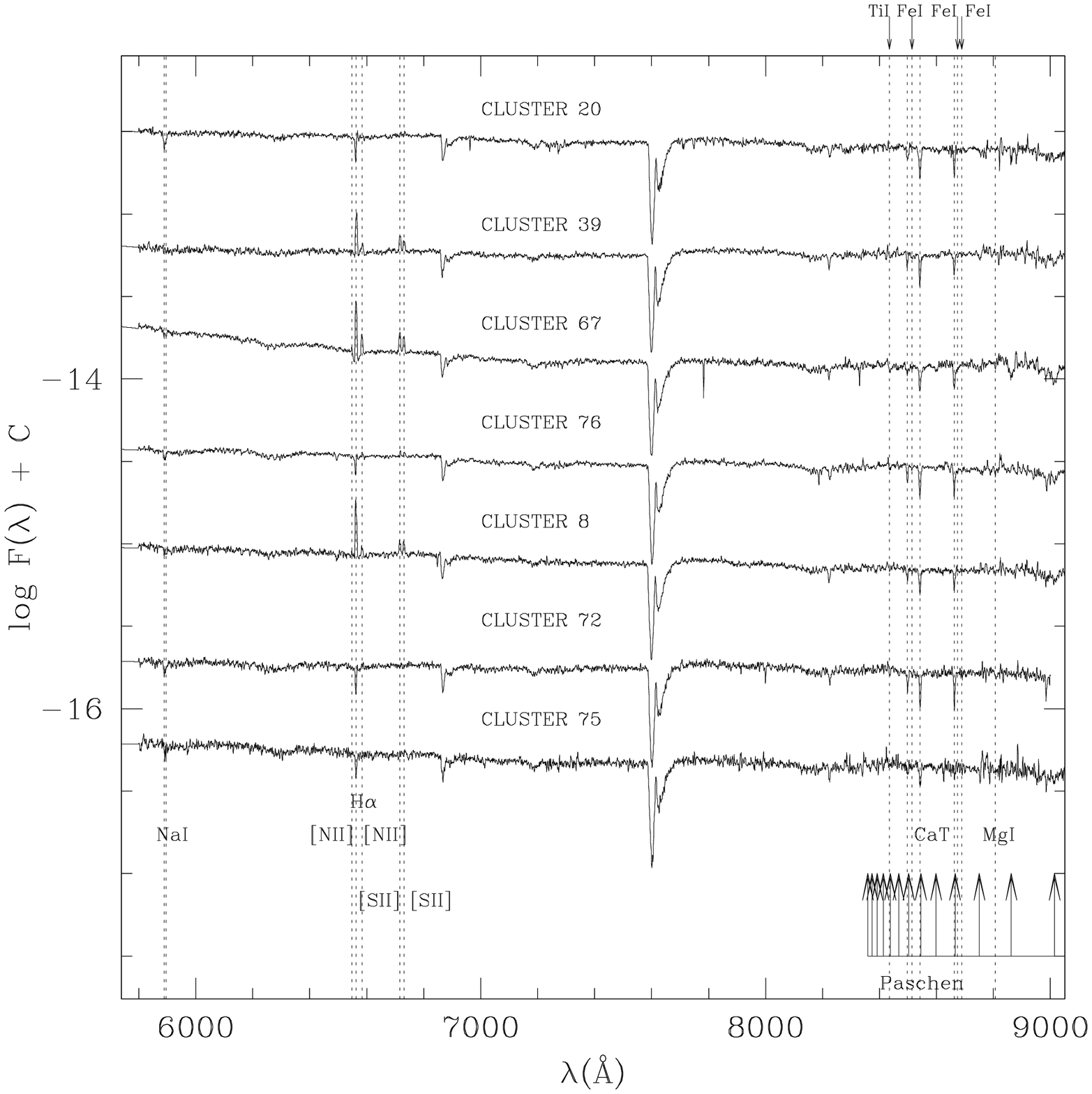}
\caption{LBT/ MODS red calibrated spectra in the rest-frame for clusters CL~20, CL~39, CL~67, CL~76, CL~8, CL~72 and CL~75 in NGC~4449. 
Indicated are the most prominent absorption and emission lines. 
\label{spectra_red_a}}
\end{figure*}

\begin{figure*}
\includegraphics[width=\textwidth]{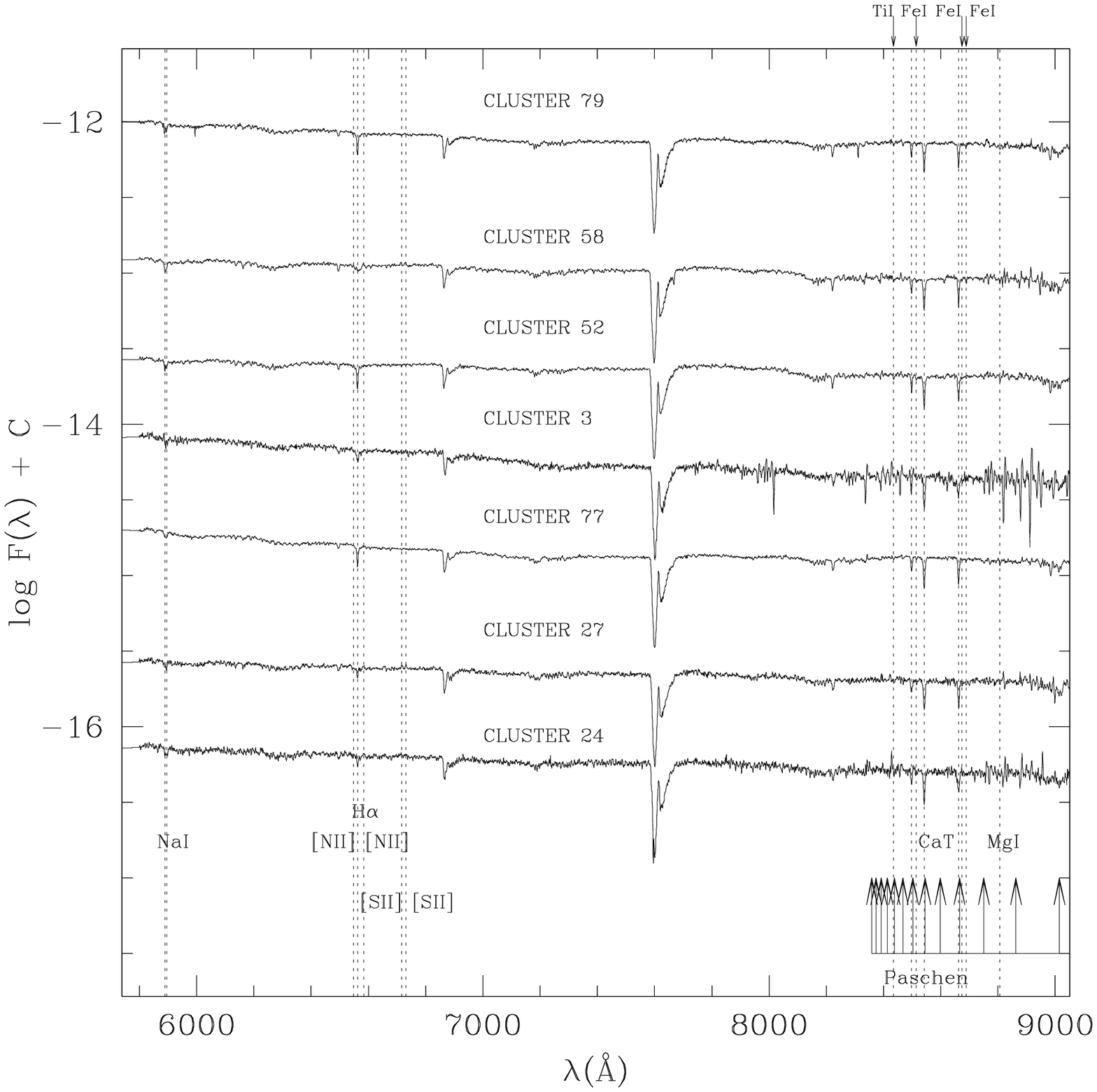}
\caption{Same as in Fig.~\ref{spectra_red_a} but for clusters  CL~79, CL~58, CL~52, CL~3, CL~77, CL~ 27 and CL~24. 
\label{spectra_red_b}}
\end{figure*}

The cluster spectra were transformed into the rest-frame system using the IRAF task {\it dpcor}. 
The final rest-frame, calibrated spectra in the blue and red MODS channels are shown in Figures~\ref{spectra_blue_a}, \ref{spectra_blue_b}, \ref{spectra_red_a} and \ref{spectra_red_b}.

\section{Narrow Band Indices} \label{index_section}

From the final calibrated spectra, we derived optical and near-infrared narrow-band indices that quantify the strength of stellar absorption lines with respect to local continua.  
More specifically, we computed the Lick indices in the optical \citep{worthey94,wo97}, and the CaII triplet index CaT* defined by  \citet{cenarro01a} and \citet{cenarro09}. 

\subsection{Lick Indices} \label{lick_section}

\begin{figure*}
\includegraphics[width=\textwidth]{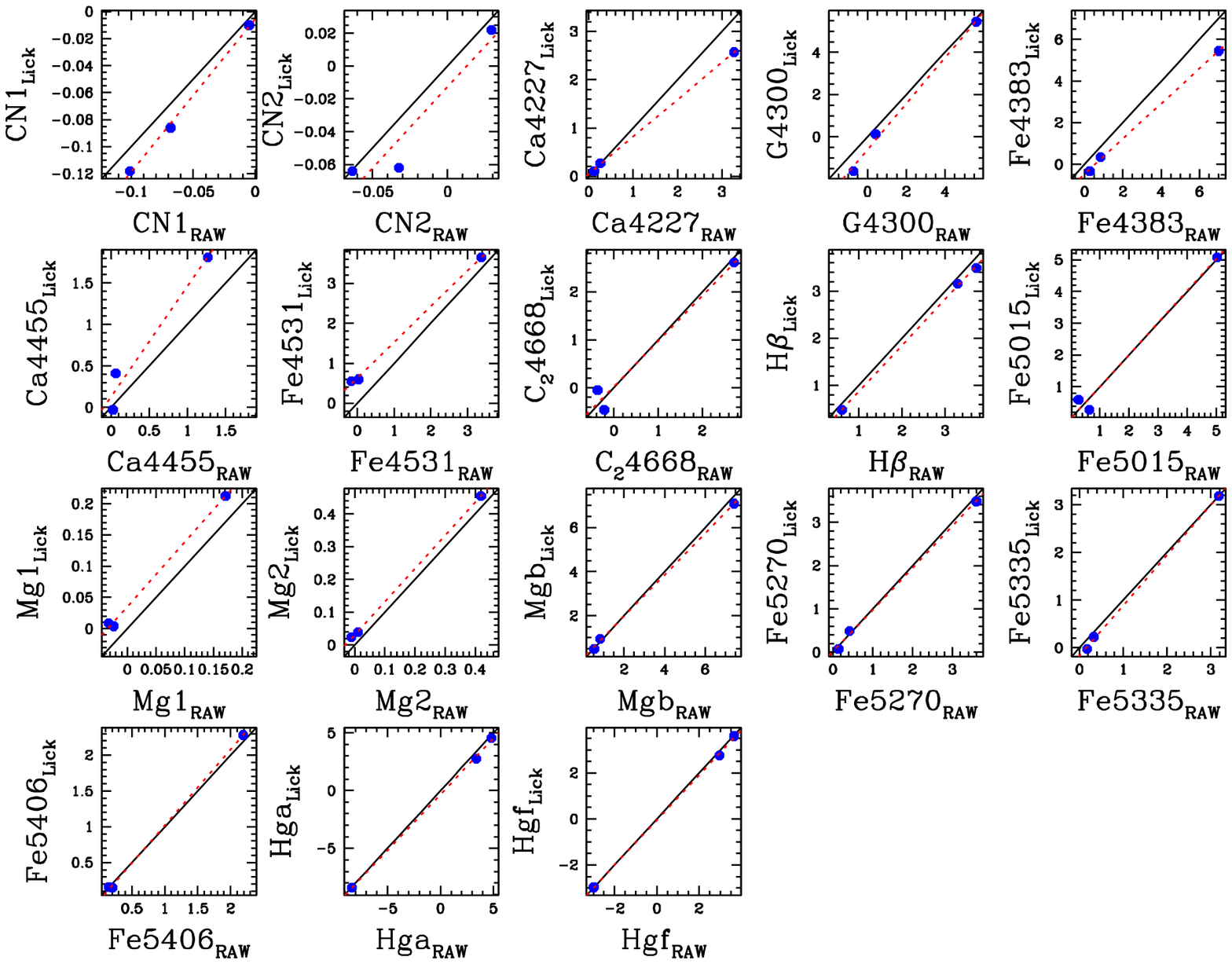}
\caption{ 
\label{lickstar} 
 Comparison between the nominal Lick index values and our measurements for the three Lick standard stars (HD~74377,  HD~84937, and HD~108177)  observed during our run.
 The solid line is the one-to-one relation, while the dotted line is the least-squares linear fit. }
\end{figure*}

For the computation of the Lick indices, the adopted procedure is the same as described in our previous papers \citep[e.g.][]{rampazzo05,dE}. 
Because of the low MODS sensitivity in the $\sim$5500-5800 \AA \ range (see Section~\ref{data_reduction}), the Lick indices Fe5709, Fe5781, NaD, 
TiO$_1$ and TiO$_2$ were not considered in our study.  The MODS spectra 
were first degraded to match the wavelength-dependent resolution of the Lick-IDS system (FWHM$\sim$8.4 \AA \ at 5400 \AA \ versus a MODS resolution of FWHM$\sim$5 \AA \ at the same wavelength) by convolution with a Gaussian kernel of variable width. Then, the indices were computed on the degraded MODS spectra using the refined passband definitions of \citet{trager98}. In order to calibrate the ``raw'' indices into the Lick-IDS system, we used the three Lick standard stars (HD~74377, HD~84937, HD~108177) observed during our run with the same 1''$\times$8'' slit used for the clusters. Following the prescription by \citet{wo97}, we degraded the standard star spectra to the Lick resolution, computed the indices, and compared our measurements with the values provided by \citet{worthey94} to derive linear transformations in the form EW$_{Lick}$ = $\alpha$  EW$_{raw}$ + $\beta$, where $EW_{raw}$ and $EW_{Lick}$ respectively are the raw and the calibrated indices,  and $\alpha$ and $\beta$ are the coefficients of the linear transformation. The transformations were derived 
only for those indices with three Lick star measurements \footnote{Tabulated index values for Lick standard stars can be found at  http://astro.wsu.edu/worthey/html/system.html}, while we discarded the remaining ones from our study.  
In the end, we were left with 18 out of the 25 Lick indices defined  by \citet{worthey94} and \citet{wo97}: CN$_1$, CN$_2$, Ca4227,  G4300, Fe4383,  Ca4455, Fe4531, C$_2$4668, 
H$\beta$, Fe5015, Mg$_1$, Mg$_2$, Mg$b$, Fe5270, Fe5335, Fe5406, H$\gamma$A, and H$\gamma$F.  
The linear transformations, which are shown in Figure~\ref{lickstar} and Table~\ref{lick_transf}, are consistent with small zero-point offsets for the majority of indices. 
No correction due to the cluster velocity dispersion was applied: in fact, while this is an important effect in the case of integrated galaxy indices, it is reasonable to neglect this correction for stellar clusters. The errors on the indices were determined through the following procedure. Starting from each cluster spectrum, we generated 1000 random modifications by adding a wavelength dependent Poissonian fluctuation corresponding to the spectral noise. Then we repeated the index computation procedure on each ``perturbed'' spectrum and derived the standard deviation. To these errors, we added in quadrature the errors on the emission corrections (see Section~\ref{em_corr}) and the scatter around the derived transformations to the Lick systems (see Table~\ref{lick_transf}).

\begin{table}
	\centering
	\caption{Linear transformations to the Lick-IDS system.}
	\label{lick_transf}
	\begin{threeparttable}
\begin{tabular}{lcccc} 
\hline
Index &  Unit & $a$  & $b$  & $rms$\\
\hline
CN$_1$ &  mag & 1.13$\pm$0.06  & -0.005$\pm$0.004 & 0.002 \\
CN$_2$ & mag & 1.0$\pm$0.3 & -0.01$\pm$0.01 & 0.01 \\
Ca4227 & \AA & 0.77$\pm$0.01 & 0.04$\pm$0.03 & 0.02 \\
G4300 & \AA & 1.09$\pm$0.08 & -0.6$\pm$0.3 & 0.2 \\
Fe4383 & \AA & 0.84$\pm$0.02 & -0.5$\pm$0.1 & 0.07 \\
Ca4455 & \AA & 1.3$\pm$0.3 & 0.1 $\pm$0.2 & 0.2 \\
Fe4531 & \AA & 0.89$\pm$0.03 & 0.63$\pm$0.06 & 0.05 \\
C$_2$4668 & \AA & 0.9$\pm$0.2 & 0.0$\pm$0.2 & 0.2 \\
H$\beta$  & \AA & 0.98$\pm$0.03 & -0.11$\pm$0.08 & 0.04 \\
Fe5015 & \AA & 1.1$\pm$0.1 & 0.0$\pm$0.4 & 0.3 \\
Mg$_1$ & mag & 1.03$\pm$0.06 & 0.035$\pm$0.006 & 0.006 \\
Mg$_2$ & mag & 1.01$\pm$0.01 &  0.032$\pm$0.004 & 0.003 \\
Mg$b$ & \AA & 0.94$\pm$0.02 & 0.1$\pm$0.1 & 0.07 \\
Fe5270 & \AA & 0.96$\pm$0.03 & 0.01$\pm$0.07 & 0.05 \\
Fe5335 & \AA & 1.05$\pm$0.03 & -0.16$\pm$0.05 & 0.04 \\
Fe5406 & \AA & 1.06$\pm$0.03 & -0.03$\pm$0.03  & 0.02 \\ 
H$\gamma$A & \AA & 0.97$\pm$0.02 & -0.3$\pm$0.1 & 0.1 \\
H$\gamma$F  & \AA & 0.98$\pm$0.02 & -0.05$\pm$0.07 & 0.07 \\
\hline
\end{tabular}
\begin{tablenotes}
\small 
\item 
$a$ and $b$ are the coefficients of the transformation $EW_{Lick} = \alpha  EW_{raw} + \beta$, where $EW_{raw}$ and $EW_{Lick}$ respectively are the raw and the calibrated indices; $rms$ is the root-mean-square deviation around the best linear fit. 
\end{tablenotes}
\end{threeparttable}
\end{table}

\subsection{Near infrared Indices} \label{cat_section}

In the near infrared, the CaII triplet ($\lambda\lambda$8498, 8542, 8662 \AA) is the strongest feature observed. Intermediate-strength atomic lines of Fe~I ($\lambda\lambda$8514, 8675, 8689, 8824 \AA), Mg~I ($\lambda$8807 \AA), and Ti~I ($\lambda$8435 \AA) are also  present. The Paschen series is apparent in stars hotter than G3 and can significantly contaminate the measurement of the CaII triplet. In order to overcome this problem,  \citet{cenarro01a} defined a new Ca~II triplet index, named CaT*, which is corrected from the contamination of the Paschen series. Adopting their definition, we computed the CaT* for our clusters in NGC~4449. In order to match the lower resolution of our spectra  (FWHM$\sim$5 \AA \  in the near infrared) to the better resolution of the stellar library used by  \citet{cenarro01a} (FWHM$\sim$1.5 \AA), we used the prescriptions by \citet{vaz03}. 
We did not consider the Mg~I and  sTiO indices defined by \citet{cenarro09}, since these turned out to be highly affected by a large spectral noise
due to the difficulty in subtracting the near-infrared sky background in our spectra (see Figs.~\ref{spectra_red_a} and ~\ref{spectra_red_b}). 


\subsection{Emission contamination} \label{em_corr}

As discussed in Section~\ref{data_reduction}, the majority of our spectra suffer contamination from the diffuse ionized gas present in NGC~4449 and the removal of the emission line contribution obtained with the background subtraction is not always satisfactory. 
Even small amounts of contamination from ionized gas can have a significant impact on the derived integrated ages \citep{serven10,concas17}: the effect  of emission is to fill in and weaken the absorption Balmer lines, resulting into apparent older integrated ages. Figures \ref{spectra_blue_a} to \ref{spectra_red_b} show the presence of residual emission in clusters CL~20, CL~39, CL~67, CL~76, CL~8, CL~72, CL~58, and CL~27. The remaining cluster spectra (CL~75, CL~79, CL~52, CL~3, CL~77,  CL~24) can reasonably be considered emission-free.  

We derived a correction to the Balmer absorption indices H$\beta$ and H$\gamma$ (the H$\delta$ indices were not used, as discussed in Section~\ref{lick_section}) for clusters CL~20,  CL~67, CL~76, CL~72, CL~58 and CL~27, while the emission was too high in clusters CL~39 and CL~8 to attempt a recovery of the true absorption lines. We used  the {\it deblend} function in the  {\it splot} IRAF task to measure the flux of the [O~III]$\lambda$5007 emission line; then, we used the relation between the [O~III] and the H$\beta$ fluxes to compute the correction to the Balmer indices. It is well known that the $F_{[O~III]}/F_{H\beta}$ ratio is subject to a large dispersion, since it strongly depends on the properties of the ionized gas. To overcome this difficulty and properly correct our data, we used the results from our study of the H~II regions in NGC~4449 \citep{annibali17}: from the six H~II regions analyzed, we obtain an average flux ratio of


\begin{equation} \label{eq1}
\frac{F_{[O~III]}}{F_{H\beta}}=3.2\pm0.7,
\end{equation}

\noindent from which the H$\beta$ emission can be derived. For H~II regions with temperature $T_e=10,000$ K and density $n_e=100 \ cm^{-3}$,  the H$\gamma$ emission is obtained from the  theoretical relations of \citet{sh95}, once the H$\beta$ flux is known:


\begin{equation}
\frac{F_{H\gamma}}{F_{H\beta}}\sim0.47.
\end{equation}

In the end, we computed the corrections to the Balmer indices by normalizing the $F_{H\gamma}$ and  $F_{H\beta}$ fluxes to the ``pseudo-continua'' defined in the Lick system.  The uncertainties on the corrections were computed by propagating both the errors on the measured fluxes and the dispersion in Eq.(\ref{eq1}). Our results  are summarized in Table~\ref{emission_tab}. We caution that the adopted emission line ratios may not be appropriate for cluster CL~58, where the emission source could be different from an 
H~II region. For instance, if the emission lines were due to the presence of a planetary nebula, a more adequate ratio would be $F_{[OIII]5007}/F_{H\beta} \sim 10$ \citep{annibali17}, resulting into lower Balmer corrections than those listed in Table~\ref{emission_tab}. For this reason, we will consider the final corrected Balmer line strengths for cluster CL~58  as upper limits.

For cluster CL~67, we could also directly derive the emission in H$\beta$  and H$\alpha$ by fitting the observed spectrum with combinations of Voigt profiles (in absorption) plus Gaussian profiles (in emission) \citep[see][for a similar case]{annibali17}. This allowed us to perform a consistency check of our procedure. From  Table~\ref{emission_tab} we get $F_{H\alpha}/F_{H\beta}=2.97\pm0.11$, in agreement with the theoretical value of $\sim$2.86 for H~II regions \citep{oster89,sh95}. Furthermore, Eq.(\ref{eq1}) provides an emission in H$\beta$ of  $(10\pm2) \times 10^{-17} erg \ s^{-1} cm^{-2}$, marginally consistent with the value of  
$(14.5\pm0.4) \times 10^{-17} erg \ s^{-1} cm^{-2}$ derived from the direct spectral fit. Notice that for cluster  CL~67 the corrections provided in Table~\ref{emission_tab} are those obtained from the direct spectral fit. The final corrected indices are provided in Table~\ref{indices_table} in the Appendix.

\begin{table*}
	\centering
\caption{Emission lines and corrections to the Balmer indices \label{emission_tab}}
	\begin{threeparttable}
	\begin{tabular}{lccccc} 
\hline
Cluster ID & F[O~III]  & FH$\beta$ & FH$\alpha$   & $\Delta EW_{H\gamma}$  & $\Delta EW_{H\beta}$\\  
 & [$erg s^{-1} cm^{-2} 10^{-17}$]  &[$erg s^{-1} cm^{-2} 10^{-17}$] & [$erg s^{-1} cm^{-2} 10^{-17}$] &  [\AA]  &  [\AA] \\
\hline  
CL~20 &  1.8$\pm$0.2   &   $-$  &   $-$ &  $0.15\pm0.04$ & $0.19\pm0.05$ \\
CL~67 &  32.7$\pm$0.2 & 14.5$\pm$0.4 &  43$\pm$1 &  $0.78\pm0.02$ & $1.67\pm0.04$ \\
CL~76 &  2.0$\pm$0.1  &   $-$  &   $-$ &    $0.08\pm0.02$ & $0.11\pm0.02$ \\ 
CL~58 & 27.9$\pm$0.2 & $-$ &   $-$ &  $0.4\pm0.1$ & $0.7\pm0.2$ \\
CL~72 & 1.3$\pm$0.1 & $-$ &   $-$ &  $0.13\pm0.03$ & $0.18\pm0.04$ \\
CL~27 & 0.77$\pm$0.08 & $-$ &   $-$ &  $0.04\pm0.01$ & $0.06\pm0.01$ \\
\hline
\end{tabular}
\begin{tablenotes}
\small 
\item 
Col.~(1): cluster name; Col. (2)-(4): measured fluxes for [O~III], H$\beta$, and H$\alpha$ in emission; Col. (5)-(6): correction to be applied to the  H$\gamma$ and H$\beta$ absorption indices.
\end{tablenotes}
\end{threeparttable}
\end{table*}

\section{Stellar population parameters for clusters in NGC~4449} \label{stpop}

\begin{figure*}
\includegraphics[width=\textwidth]{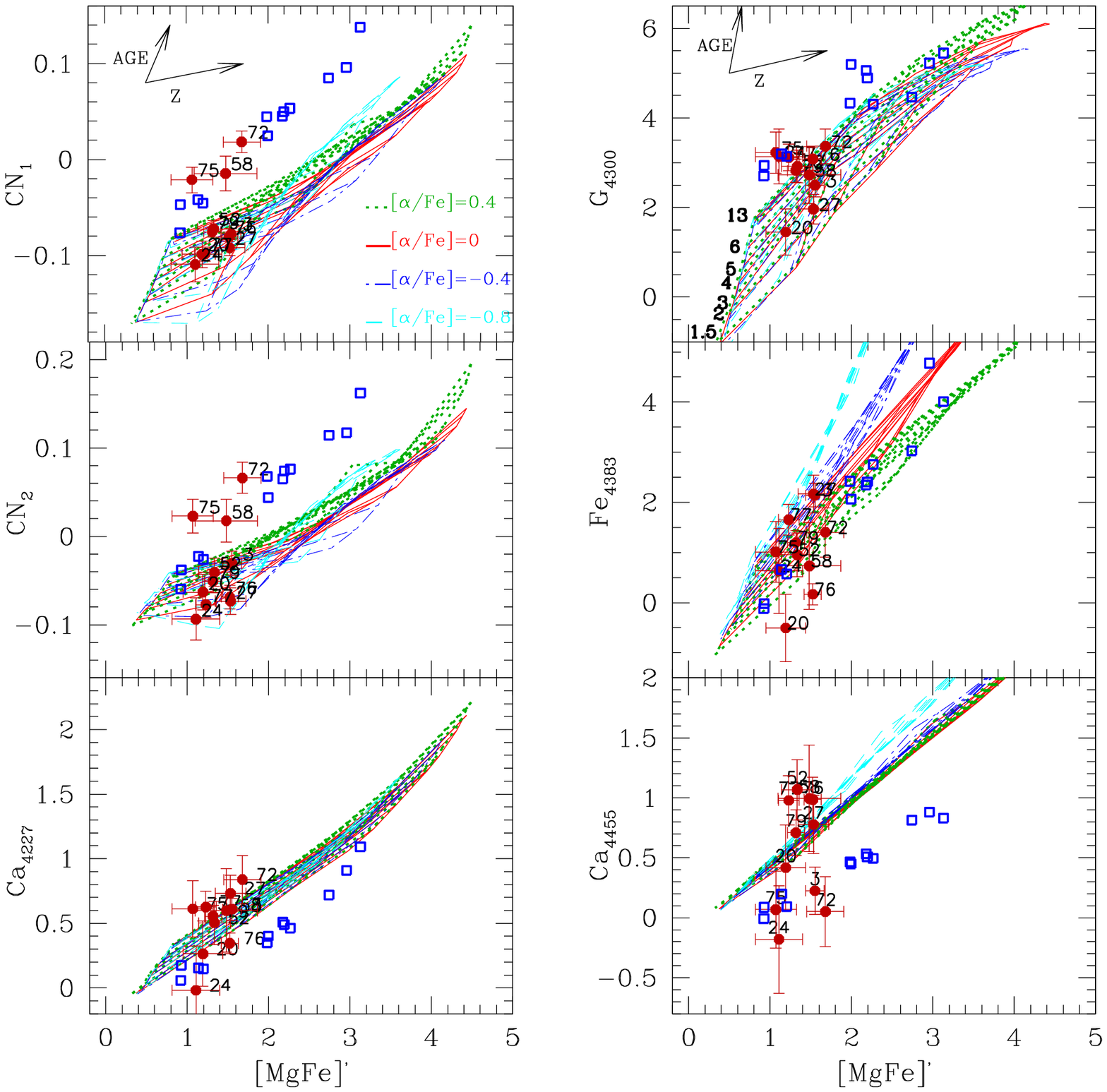}
\caption{Comparison of cluster Lick indices (CN$_1$, CN$_2$, Ca~4227, G~4300, Fe~4383, Ca~4455 versus [MgFe]$^{'}$) with our simple stellar population models. The solid red points are the NGC~4449 clusters from this work, while the open blue squares are  the MW halo and bulge globular clusters from \citet{puzia02}. The displayed models are for metallicities Z$=$0.0004, 0.004, 0.008, 0.02, 0.005, ages from 1 to 13 Gyr, and  [$\alpha$/Fe] = $-0.8, -0.4, 0.$, and 0.4. }  
\label{mod1}
\end{figure*}

\begin{figure*}
\includegraphics[width=\textwidth]{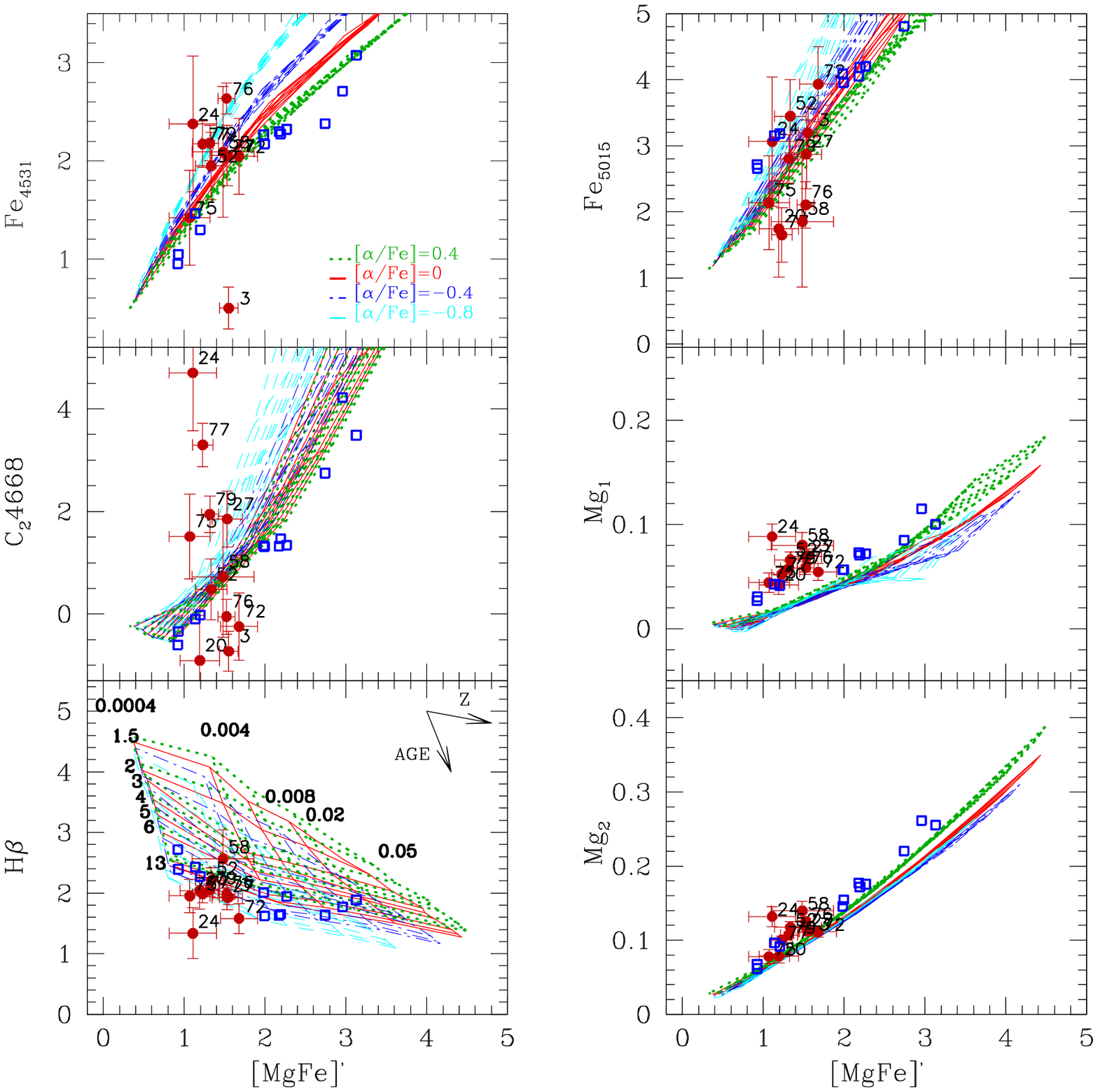}
\caption{Same as in Fig.~\ref{mod1} but for the Lick indices Fe~4351, C$_2$4668, H$\beta$, Fe~5015, Mg$_1$, and Mg$_2$. The metallicity and age labels (Z and Gyr, respectively) refer to the [$\alpha$/Fe]$=$0 models.} 
\label{mod2}
\end{figure*}

\begin{figure*}
\includegraphics[width=\textwidth]{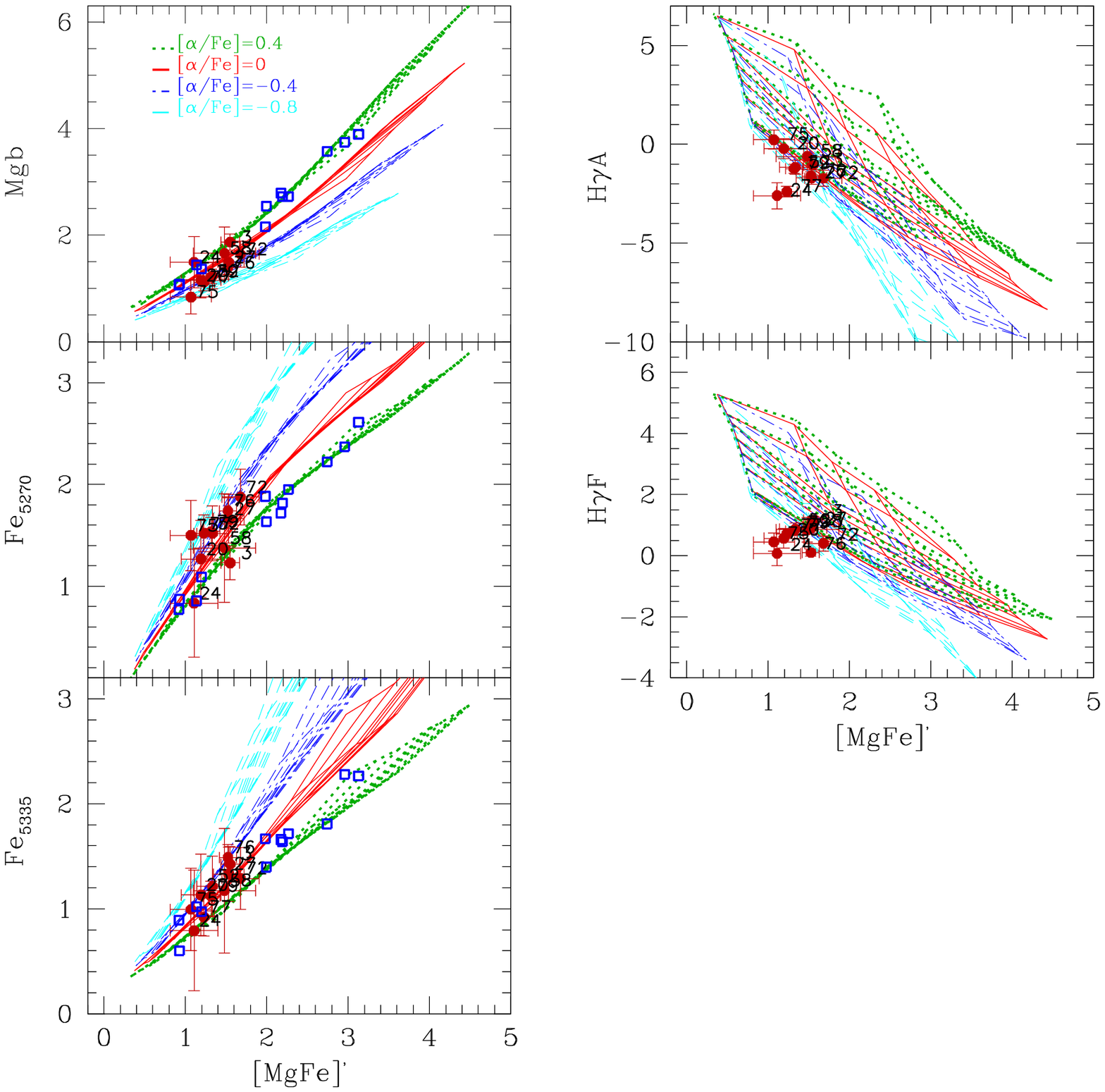}
\caption{Same as in Fig.~\ref{mod1} but for the Lick indices Mg$b$, Fe~5270, Fe~5335, H$\gamma$A, and H$\gamma$F. } 
\label{mod3}
\end{figure*}

\begin{figure*}
\includegraphics[width=\textwidth]{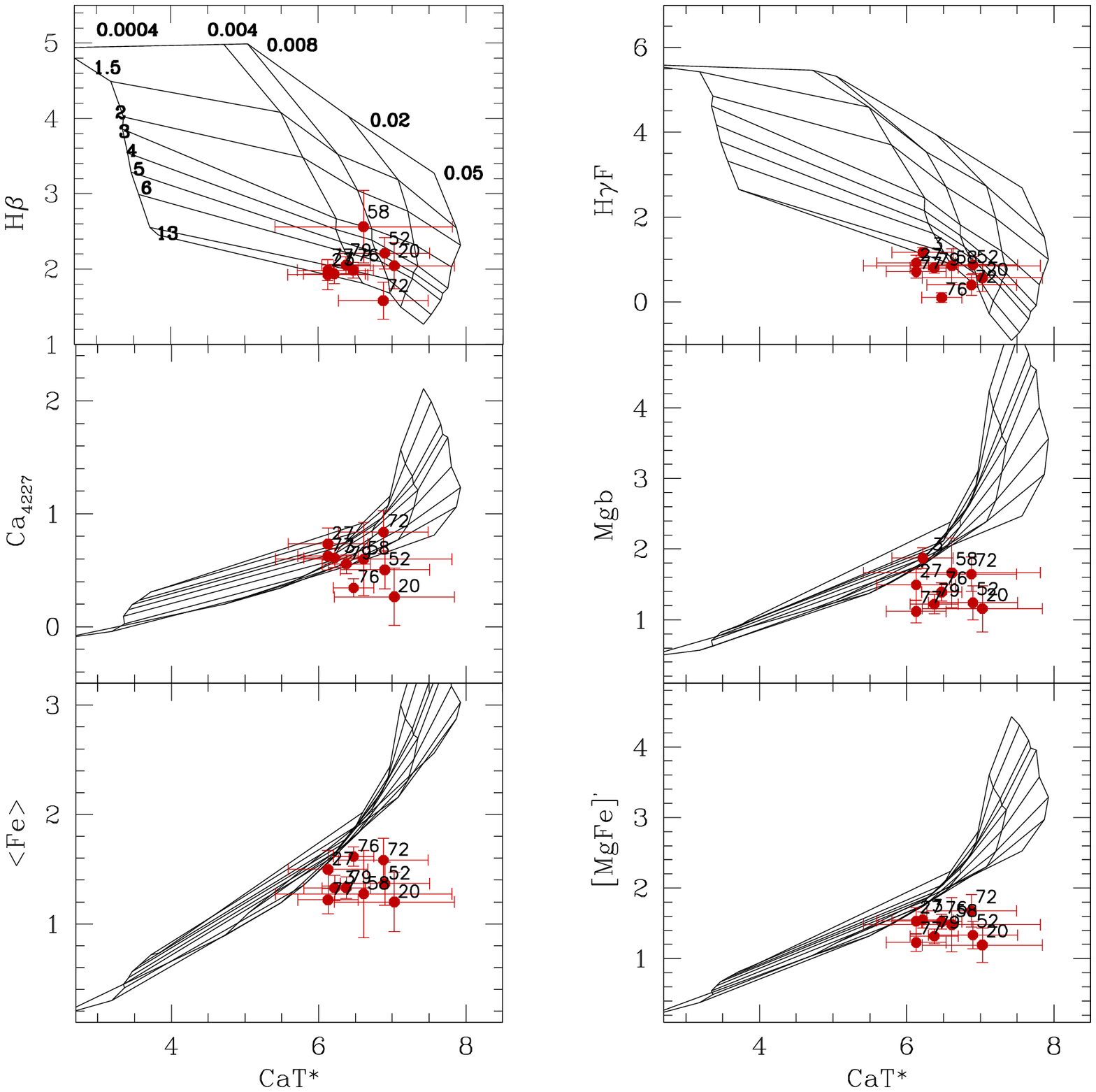}
\caption{Cluster Lick indices versus the near infrared CaT* index. The data are for the NGC~4449 clusters, while the models are our ``standard''  simple stellar populations for 
 Z$=$0.0004, 0.004, 0.008, 0.02, 0.005, and ages from 1 to 13 Gyr. Notice that the ``standard'' models are calibrated on Galactic stars and thus reflect the MW 
chemical composition ratios. }
\label{mod4}
\end{figure*}

In Figures~\ref{mod1} to ~\ref{mod4} we compare the  indices derived for the NGC~4449's clusters with our set of  simple stellar population (SSP) models \citep{annibali07,dE}.  
 The models are based on the Padova SSPs  \citep[][and references therein]{bressan94}, and on the fitting functions (FFs) of \citet{worthey94} and \citet{wo97}. 
 The SSPs were computed for different element abundance ratios, where the departure from the solar-scaled composition is based on the index responses by \citet{korn05}.  In the models displayed in Figs.~\ref{mod1} to~\ref{mod3}, the elements O, Ne, Na, Mg, Si, S, Ca, and Ti are assigned to the $\alpha$-element group, Cr, Mn, Fe, Co, Ni, Cu, and Zn to the Fe-peak group. We assume that the elements within one group are enhanced/depressed by the same factor; the $\alpha$ and  Fe-peak elements are respectively enhanced and depressed in the [$\alpha$/Fe]$>$0 models, while the opposite holds for the [$\alpha$/Fe]$<$0 models. Other elements, such as N and C, are left untouched and scale with the solar composition.  We caution that all these assumptions are an over-simplification: in fact, recent studies suggest that Mg possibly behaves differently from the other $\alpha$'s \citep[see][their Section 4.3]{pancino17} and it is well known that Galactic globular clusters exhibit an anti-correlation between Na and O, and between C and N \citep[e.g.][]{carretta06};
an anti-correlation between Mg and Al has also been observed for the most metal-poor and/or most massive GCs \citep{pancino17}.
Furthermore, notice that the stellar evolutionary tracks adopted in our models have solar-scaled chemical compositions \citep{fagottoa,fagottob} and that only the effect due to element abundance variations in the model atmospheres, as quantified by the \citet{korn05} response functions, is included.

 In Figures~\ref{mod1} to ~\ref{mod3}, we plot the Lick indices for our clusters against the metallicity sensitive [MgFe]$^{'}$ index\footnote{Defined as $[MgFe]^{'}=\sqrt{Mgb \times ( 0.72\cdot Fe5270 + 0.28 \cdot Fe5335)}$.}, which presents the advantage of being mostly insensitive to Mg/Fe variations \citep{gonza93,tmb03}.  
 For comparison, we also show the indices derived by \citet{puzia02} for Galactic clusters in the halo and in the bulge. In Fig.~\ref{mod4}, NGC~4449' s clusters are plotted instead on optical-near infrared planes, with a Lick index versus CaT*;  here the models have solar-scaled composition since, to our knowledge, specific responses of the CaT* index to individual element abundance variations  have not been computed yet. 

 \subsection{Discussion on individual indices} \label{ind_disc}
 
The {\bf CN$_1$} and  {\bf CN$_2$} indices measure the strength of the CN absorption band at $\sim$4150 \AA, and exhibit a strong positive response to the abundances of C and N.  
On the other hand, they are almost insensitive to [$\alpha$/Fe] variations.  
In Fig.\ref{mod1}, the bulk of clusters in NGC~4449 nicely overlaps the SSP models. The MW clusters, instead,  exhibit a significant offset from the models, a behaviour that has been explained as due to a significant nitrogen enhancement \citep[e.g.][]{origlia02,puzia02,tmb03}. Three clusters in our sample, namely CL~72, CL~75, and CL~58, occupy the same high CN strength region of the Galactic globular clusters.
In order to match the \citet{puzia02} Galactic globular cluster data with their SSP models, \citet{tmb03} needed  to assume  [N/$\alpha$]$=0$.5 (i.e. a factor 3 enhancement in nitrogen with respect to the $\alpha$-elements).  
Notice that the CN features of the Galactic bulge are instead perfectly reproduced by models without N enhancement \citep{tmb03}. We will come back to the problem of the N abundance in NGC~4449' s clusters later in this paper. 

 
The {\bf Ca~4227} index is dominated by the Ca$\lambda$4227 absorption line, strongly correlates with metallicity and, interestingly, is the only index besides CN$_1$ and  CN$_2$ to be affected by changes in the C and N abundances (i.e., it anti-correlates with CN). While the MW GCs are shifted to lower Ca~4227 values compared to the models (a discrepancy that according to \citet{tmb03} can be reconciled assuming [N/$\alpha$]$=0$.5 models) the NGC~4449 clusters agree pretty well with the SSPs. This behaviour is in agreement with the good match between data and models in the {\bf CN$_1$} and  {\bf CN$_2$} indices. 
The {\bf Ca~4455} index, on the other hand, is not a useful abundance indicator. In fact, despite its name, it is a blend of many elements, and it has been shown to be actually insensitive to Ca \citep{tb95,korn05}. It exhibits a very small dependence on the [$\alpha$/Fe] ratio. Our data are highly scattered around the models, with quite large errors on the index measurements. 
Hereafter, we will not consider this index  for our cluster stellar population analysis.

The {\bf G~4300} index measures the strength of the G band and is highly sensitive to the CH abundance. It exhibits a dependence on both age and metallicity, and no significant sensitivity to the [$\alpha$/Fe] ratio, so that the  G4300 vs. [MgFe]$^{'}$ plane can be used to separate  age and metallicity to some extent. 
According to Fig.~\ref{mod1}, our clusters have metallicities Z$\lesssim$0.004 and ages in the $\sim3 - 13$  Gyr range.  However, \citet{tmb03} noticed that the calibration of this index versus Galactic globular cluster data is not convincing, therefore it is not clear if G4300 can provide reliable results on the stellar population properties. 

The {\bf C$_2$4668} index, formerly called Fe4668, is slightly sensitive to Fe and most sensitive to C abundance. It  exhibits a low negative correlation with the [$\alpha$/Fe] ratio. The NGC~4449 clusters data are highly scattered in the C$_2$4668 vs [MgFe]$^{'}$ plane, possibly as a consequence of the large error  in the C$_2$4668 measurement.
As discussed by \citet{tmb03}, this index is not well calibrated against Galactic globular cluster data and therefore is not well suited for element abundance studies. 

Among the indices that quantify the strength of Fe lines, {\bf Fe~5270} and {\bf Fe~5335} are the ones most sensitive to [$\alpha$/Fe] variations. 
The other {\bf Fe~4383}, {\bf Fe~4531}, and {\bf Fe~5015} indices contain blends of many metallic lines other than Fe (e.g. Ti~I, Ti~II, Ni~I) and may be less straightforward than Fe~5270 and Fe~5335  to interpret. Fig.~\ref{mod3} shows that models of different [$\alpha$/Fe] ratios are very well separated in the Fe~5270, Fe~5335 vs. [MgFe]$^{'}$ planes, with the 
NGC~4449' s clusters preferentially located in the region of [$\alpha$/Fe]$<$0.  The same behaviour is observed in the {\bf Mg$b$} vs. {\bf [MgFe]$^{'}$} plane. The Mg$b$ index quantifies the strength of the Mg~I~b triplet at  5167, 5173, and 5184 \AA, and is the most sensitive index to [$\alpha$/Fe] variations among the Mg lines. Indeed, the {\bf Mg$_1$} and {\bf Mg$_2$} dependence on [$\alpha$/Fe] is quite modest, in particular at low metallicities. The Mg$_2$ index is centered on the Mg~I~b feature, as the Mg$b$, while the Mg$_1$ samples MgH, Fe~I, and Ni~I lines at $\sim$4930 \AA. Both  Mg$_1$ and Mg$_2$  pseudo-continua are defined on a $\sim$400 \AA \ baseline, much larger than the typical $\sim$100 \AA \ baseline of the other Lick indices, implying that Mg$_1$ and Mg$_2$ are potentially affected by uncertainties on the relative flux calibration. This could explain the mismatch observed between our cluster data and the models for both the indices: in fact, as discussed in Section~\ref{data_reduction}, our observations were highly affected by differential flux losses due to atmospheric differential refraction and it is possible that our correction, based on a few photometric bands, was not able to perfectly correct for this effect.

\begin{figure*}
\includegraphics[width=\textwidth]{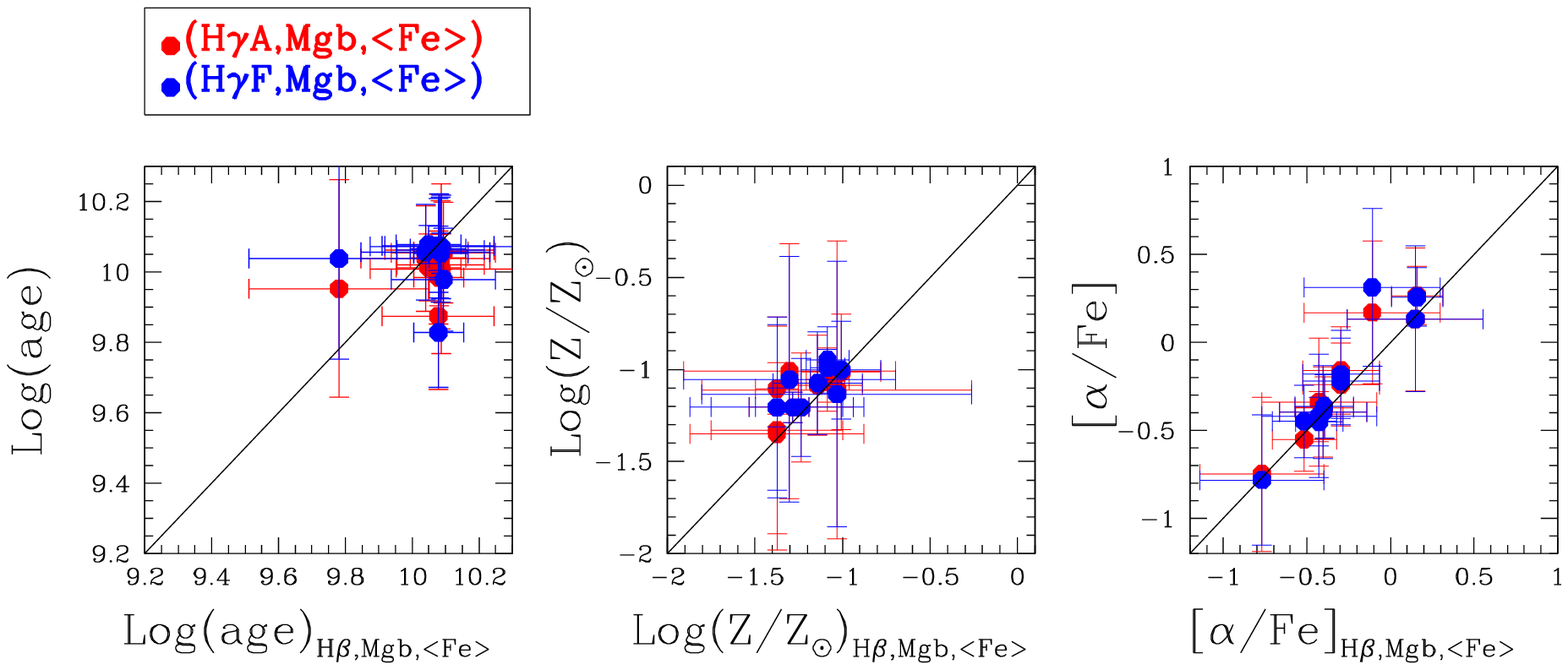}
\caption{Ages, metallicities, and [$\alpha$/Fe] ratios derived for clusters CL~20, CL~76, CL~72, CL~75, CL~79, CL~58, CL~52, CL~3, CL~77, CL~27 and CL~24 in NGC~4449 using different combinations of index triplets. 
The stellar population parameters obtained with the (H$\beta$, Mg$b$, $<Fe>$) triplet are on the x axes, while the results from the other triplets, as indicated in the legend, are on the y axes. The solid line is the one-to-one relation. Ages are in units of $\log_{10}(age [yr])$. The adopted solar metallicity is  $Z_{\odot}=0.018$.}
\label{triplets}
\end{figure*}

Finally, the {\bf H$\beta$}, {\bf H$\gamma$A}, and {\bf H$\gamma$F} vs. [MgFe]$^{'}$  planes provide the best separation between age and total metallicity. The H$\beta$ index is poorly affected  
by [$\alpha$/Fe] variations, but suffers possible contamination from Balmer emission. On the other hand, the H$\gamma$ indices are less affected by emission than H$\beta$, but present a marked dependence on [$\alpha$/Fe] due to the presence of several Fe lines in their pseudo-continua. 
We recall that our Balmer indices  have been corrected for the presence of residual emission lines on the cluster spectra; the two only exceptions are clusters CL~39 and CL~8,
not shown here, whose emission was too strong to attempt a recovery of the true absorption strengths. Cluster CL~24 was one of those with no visible emission lines in
the final subtracted spectra, nevertheless we notice that it falls slightly below the oldest models in all the H$\beta$, H$\gamma$A, H$\gamma$F vs. [MgFe]$^{'}$  planes. 
All the other clusters fall within the model grid and are consistent with Z$\lesssim$0.004 (or  Z$\lesssim$0.008, if we consider the models with [$\alpha$/Fe]$=-0.8$) and 4 Gyr $\lesssim$age$\lesssim$ 13 Gyr. A more quantitative analysis of the stellar population parameters will be performed in the next section. 
Cluster CL~67, with Balmer indices as high as $H\beta=6.4\pm0.6$, $H\gamma A=10.1\pm0.7$, and $H\gamma F=7.3\pm0.4$, falls outside the displayed index ranges, and it is a few hundreds Myr old.

\begin{table*}
	\centering
\caption{Derived stellar population parameters for clusters in NGC~4449 \label{agezafe_tab}}
\begin{tabular}{lccccc} 
\hline
Cluster & Age [Gyr] & $\log(Z/Z_{\odot})$  & [$\alpha$/Fe]  & [Fe/H]  & [N/$\alpha$]  \\
\hline
 CL~20  &  11$\pm$5 & $-1.3\pm0.5$ & $-0.4\pm0.3$ & $-1.0 \pm 0.6$  &   $<-0.5$\\  
 CL~76  &  11$\pm$2 & $-1.0\pm0.1$ & $-0.4\pm0.1$ & $-0.7\pm0.2$  &  $-0.4\pm0.2$  \\  
CL~72  &  11$\pm$4 & $-1.0\pm0.3$ & $-0.2\pm0.2$ & $-0.8\pm0.4$  &  $>0.5$ \\  
CL~75  &  10$\pm$4 & $-1.3\pm0.5$ & $-0.8\pm0.4$ & $-0.7\pm0.6$  &  $>0.5$ \\  
 CL~79  &  11$\pm$2 & $-1.2\pm0.1$ & $-0.4\pm0.2$ &  $-0.9\pm0.2$  &   $-0.2\pm0.2$ \\  
CL~58  &   $\ge$9 & $\le-1.1$ & 0.1$\pm$0.4  & $\le-1.2$  & $0.5\pm0.4$\\  
CL~52  &    11$\pm$4 & $-1.2\pm0.3$ & $-0.4\pm0.3$ & $-0.9\pm0.4$ &  $-0.1\pm0.1$  \\  
CL~3  &    9$\pm$2 &  $-1.0\pm0.2$ & 0.2$\pm$0.2  &  $-1.2\pm0.2$  &    $-0.3\pm0.2$   \\  
CL~77  &    12$\pm$2 &  $-1.2\pm0.1$ & $-0.5\pm0.2$ & $-0.8\pm0.2$   &  $<-0.5$\\  
CL~27  &    11$\pm$4 &  $-1.1\pm0.3$ & $-0.3\pm0.2$  & $-0.9\pm0.3$ & $-0.5\pm0.2$ \\  
CL~24  &    12$\pm$4 &  $-1.1\pm0.6$ & $0.1\pm0.4$  & $-1.2\pm0.8$ & $<-0.5$  \\  
\hline
\end{tabular}
\end{table*}

In Fig.~\ref{mod4} we plotted some key Lick indices versus the near infrared {\bf CaT*} (Ca II triplet) index. We did not consider  cluster CL~75 whose near-infrared spectrum was too noisy due to a non-optimal sky background subtraction. As previously noticed, no models with variable [$\alpha$/Fe] ratios have been computed for the near-infrared indices and therefore the displayed SSPs refer to the base-model \footnote{The base model uses fitting functions \citep{cenarro02} calibrated on Galactic stars, and therefore reflects the MW chemical composition: [$\alpha$/Fe]$=$0 at solar metallicities, and [$\alpha$/Fe]$>0$ at sub-solar metallicities.}.  Qualitatively, the CaT* index provides results that are 
consistent with the Lick indices, with cluster total metallicities of  Z$\lesssim$0.008. The H$\beta$ vs CaT* plane seems to be a viable alternative to the  H$\beta$ vs [MgFe]$^{'}$ diagram 
to disentangle age and metallicity effects. Although a discussion on the cluster element abundance ratios is not possible due to the lack of models different from the base one, we immediately notice the significant mismatch between data and models in the Mg$b$ vs. CaT* plane, confirming the low [$\alpha$/Fe] ratios for our clusters.




\subsection{Ages, metallicities and [$\alpha$/Fe] ratios} \label{agezafe}

If we exclude the clusters with very strong emission contamination (i.e., clusters  CL~39, CL~67, and CL~8) 
from our sample, we are left with a sub-sample of 11 clusters, namely CL~20, CL~76, CL~72, CL~75, CL~79, CL~58, CL~52, CL~3, CL~77, CL~27 and CL~24. 
For these clusters, we derived ages, metallicities, and [$\alpha$/Fe] ratios using the algorithm described in \citet{annibali07}. In brief, each SSP model of given age (t), metallicity (Z), and [$\alpha$/Fe] ratio ($\alpha$) univocally corresponds to a point in a  three-dimensional space defined  by an index triplet. For each measured index triplet, we  compute the {\it likelihood} that the generic (t,Z,$\alpha$) model be the solution to that data point. 
This procedure provides a {\it likelihood} map in the three-dimensional (t,Z,$\alpha$) space, and allows us to derive the ``most probable'' solution with  its associated uncertainty. 

Following our discussion in Section~\ref{ind_disc}, we selected in the first place the following indices for our stellar population study:
H$\beta$, H$\gamma$A, H$\gamma$F, Mg$b$, Fe~5270, and Fe~5335.  We then considered the following index triplet combinations, composed of an age-sensitive index and two metallicity-sensitive indices: (H$\beta$, Mg$b$, $<Fe>$), (H$\gamma$A, Mg$b$, $<Fe>$), (H$\gamma$F, Mg$b$, $<Fe>$), where $<Fe>=0.5\times(Fe~5270+Fe~5335)$.
The CN$_1$, CN$_2$, and Ca~4227 indices,  potentially affected by CN variations, were analyzed in a second step. 
For each triplet, we derived ages, total metallicities Z, and [$\alpha$/Fe] ratios using our SSPs \citep{annibali07,dE} and the algorithm described at the beginning of this section. 
The results are displayed in Fig.~\ref{triplets}, where we plot the stellar population parameters obtained with the three different triplets.  
Fig.~\ref{triplets}, left panel, shows that the typical errors on the derived ages are quite large; the results obtained adopting different Balmer indices 
are highly scattered, although typically consistent with each other given the large errors on the ages. 
This large age uncertainty is the natural consequence of the progressively reduced age-sensitivity of the Balmer absorption lines with increasing age: for instance, a $\sim$0.2 \AA  \ difference on the H$\beta$ index (which is the typical error for our clusters) corresponds to several Gyrs  difference at old ages.  On the other hand, the metallicity and the [$\alpha$/Fe] values obtained with the different triplets are in quite good agreement, despite the age-metallicity degeneracy; this is due to the high capability of the Balmer $+$ Mgb $+$ $<$Fe$>$ diagnostic in separating age, metallicity and $\alpha$/Fe affects \citep[e.g][]{tmb03}.

In the end, we derived final ages, metallicities and [$\alpha$/Fe] ratios by averaging the results from the three different triplets, and computed the errors by propagating the 
uncertainties on the stellar population parameters. 
Table~\ref{agezafe_tab}, where we summarize our results, shows that the majority of the  analyzed clusters are typically old 
($\gsim$9 Gyr).
The total metallicities are highly sub-solar, with $-1.3 \lesssim \log(Z/Z_{\odot}) \lesssim -1.0$; the majority of clusters have sub-solar [$\alpha$/Fe] ratios (in the range $-0.2$--$0.5$), and only three clusters exhibit slightly super-solar [$\alpha$/Fe] ratios. 
[Fe/H] values were computed from the total metallicity Z and the [$\alpha$/Fe] ratio through to the formula: 

\begin{equation}
[Fe/H] = \log(Z/Z_{\odot}) + \log(f_{Fe}) 
\end{equation}

\noindent where we have assumed $Z_{\odot}=0.018$, and $f_{Fe}$ is the enhancement/depression factor of the Fe abundance in the models \citep[see][for details]{annibali07}.

\begin{figure*}
\includegraphics[width=\textwidth]{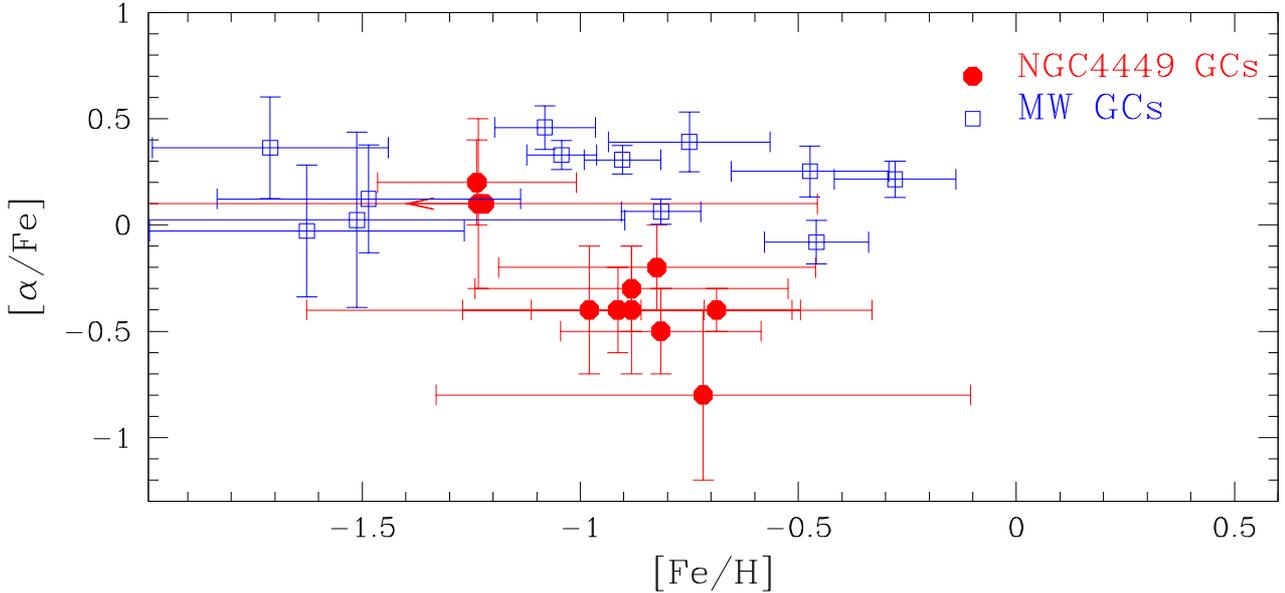}
\caption{Distribution of clusters in NGC~4449 and in the MW in the [$\alpha$/Fe] versus [Fe/H] plane. The displayed MW clusters belong to the sample of \citet{puzia02} and are both halo and bulge globular clusters. The stellar population parameters of Galactic clusters have been obtained by reprocessing the Lick indices provided by  \citet{puzia02} through our models for a self-consistent comparison with the NGC~4449 results.} \label{afe_feh}
\end{figure*}

NGC~4449' s clusters are displayed in the [$\alpha$/Fe] versus [Fe/H] plane in Fig.~\ref{afe_feh}, together with Milky Way halo and bulge clusters from \citet{puzia02}.
For a self-consistent comparison, the Lick indices provided by \citet{puzia02} were re-processed through our algorithm and models. 
The difference between clusters in NGC~4449 and in the  Milky Way is striking in this plane: while Galactic globular clusters exhibit a flat distribution with solar or super-solar [$\alpha$/Fe] ratios at all metallicities, the clusters in NGC~4449 display a trend, with slightly super-solar [$\alpha$/Fe] ratios at the lowest  cluster metallicities, and highly sub-solar 
[$\alpha$/Fe] values at $[Fe/H]>-1$. Furthermore, the $[Fe/H]$ range of NGC~4449' s clusters is between $\sim-1.2$  and $-0.7$, higher than MW halo GCs  and comparable to MW bulge clusters.


\subsection{Nitrogen and Carbon} \label{nitrogen_carbon}

\begin{figure*}
\includegraphics[width=\textwidth]{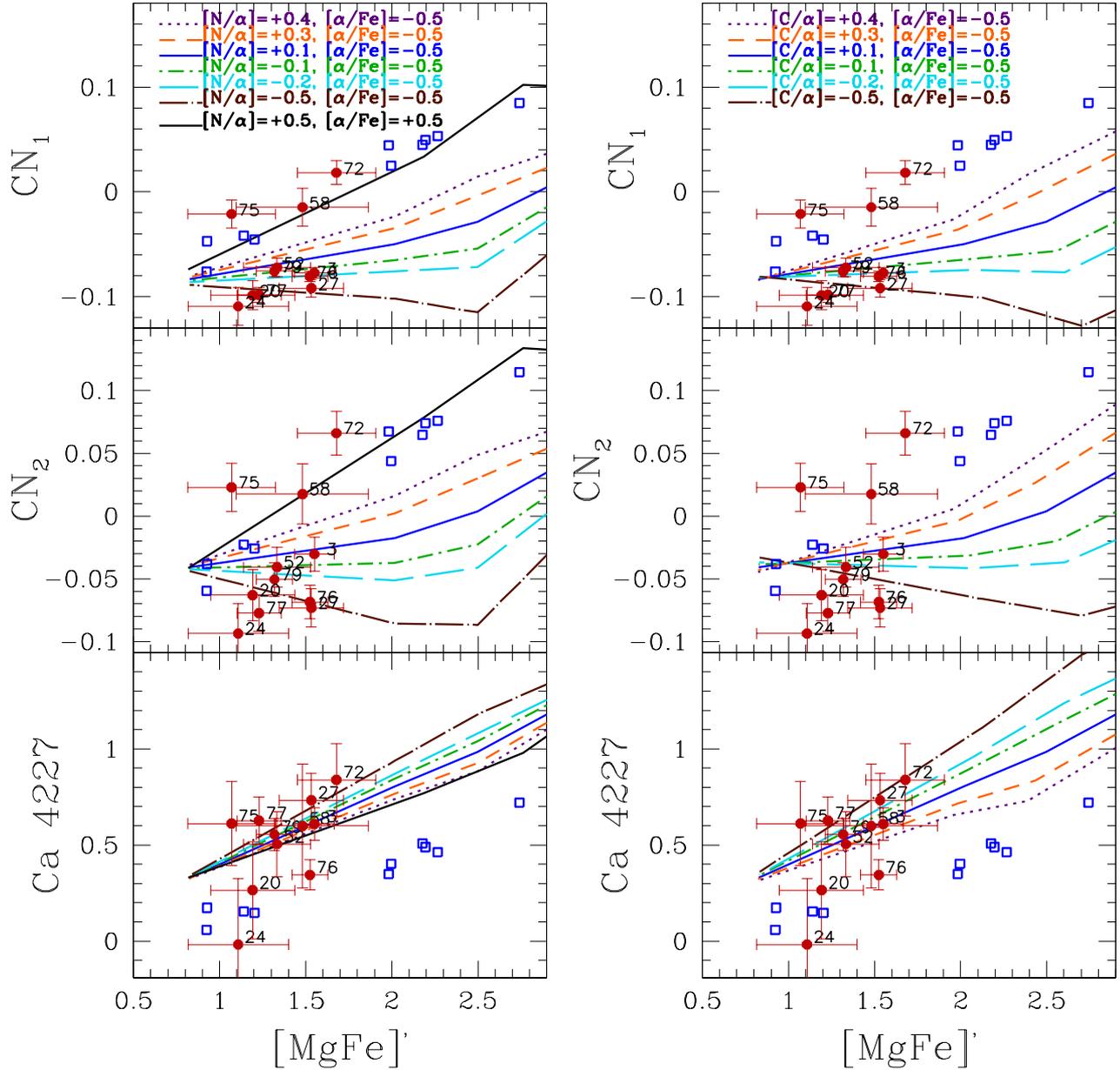}
\caption{Distribution of NGC~4449' s and MW clusters from \citet{puzia02} (full red points and open blue squares, respectively) in the CN$_1$, CN$_2$, and Ca~4227 vs. 
 [MgFe]$^{'}$  planes. Overplotted are 11 Gyr old models for different combinations of [$\alpha$/Fe] and [N/$\alpha$].} \label{fig_nvar}
\end{figure*}

The CN$_1$, CN$_2$, and Ca~4427 indices, which are affected by CN abundance variations, can be used as diagnostics to study the nitrogen and carbon composition in our clusters. Fig.~\ref{mod1}, left panels, shows a remarkable difference in CN$_1$, CN$_2$, and Ca~4427 between the bulk of clusters in NGC~4449 and Galactic globular clusters. In order to quantify this behaviour in terms of N and C abundance differences, 
we computed additional SSP models where we increased or depressed the abundance of C or N with respect to the solar-scaled composition. This resulted into models with different combinations of ([$\alpha$/Fe], [N/$\alpha$]) or ([$\alpha$/Fe], [C/$\alpha$]), shown in Fig.~\ref{fig_nvar} together with  the NGC~4449 and MW cluster data. The plotted models correspond to an age of 11 Gyr and [$\alpha$/Fe]$=-0.5$ to reflect the typical stellar population parameters derived for the NGC~4449's clusters (see Table~\ref{agezafe_tab}) and have [N/$\alpha$] (or [C/$\alpha$]) values from $-0.5$ to  $+0.5$. We also plotted a 11 Gyr model with [$\alpha$/Fe]$=+0.5$ and [N/$\alpha$]$=+0.5$ to account for the typical abundance ratios derived in Galactic globular clusters from integrated-light absorption features \citep[e.g.,][]{tmb03}.  We notice that while the CN indices of Galactic globular clusters are well reproduced by N-enhanced models, the majority of clusters in NGC~4449,  with the exception of clusters CL~72, CL~75, and CL~58 that are more compatible with the Galactic ones, require solar or sub-solar [N/$\alpha$] (or [C/$\alpha$]) ratios to match the models. The Ca~4227 index is less affected than CN$_1$ and CN$_2$ by CN abundance variations and it is therefore less useful to  characterize  the chemical path of C and N. The Ca~4227 indices measured by \citet{puzia02}  for Galactic globular clusters are not well reproduced by the 
[$\alpha$/Fe]$=+0.5$, [N/$\alpha$]$=+0.5$ models and show an offset as large as  $\sim-0.3$ \AA. 

The CN$_1$, CN$_2$, and Ca~4427 indices alone do not allow us to break the degeneracy between C and N abundance variations. 
The C$_2$4668 and Mg$_1$ indices potentially offer a way out since they are much more sensitive to C than to N, but for  our clusters they  exhibit large errors or are not well calibrated, which makes them not very useful for our study. \citet{tmb03} showed that C-enhanced models could reproduce the large CN$_1$ and CN$_2$ indices observed in  
MW globular clusters, but failed in reproducing C$_2$4668 and Mg$_1$;  therefore they  concluded that a nitrogen rather than a carbon enhancement was more likely in Galactic globular clusters. Building on these results, we computed [N/$\alpha$]  values for the NGC~4449 clusters through comparison of the CN$_1$ and CN$_2$ indices against models where N is enhanced/depressed and C is left unchanged to the solar-scaled composition. We input into our algorithm,  described in Section~\ref{agezafe}, the age, metallicity, and [$\alpha$/Fe] values listed in Table~\ref{agezafe_tab}, derived two different [N/$\alpha$] ratios respectively from the 
(CN$_1$, Mgb, $<$Fe$>$) and (CN$_2$, Mgb, $<$Fe$>$) triplets, and then averaged the results. The mean [N/$\alpha$] values, listed in Col.~6 of Table~\ref{agezafe_tab}, are  between $-0.5$ and $-0.1$ for all clusters but   CL~72, CL~75, and CL~58, which exhibit highly super-solar  [N/$\alpha$] ratios. 

\section{A planetary nebula within cluster CL~58?} \label{cl58_section}

\begin{figure*}
\includegraphics[width=\textwidth]{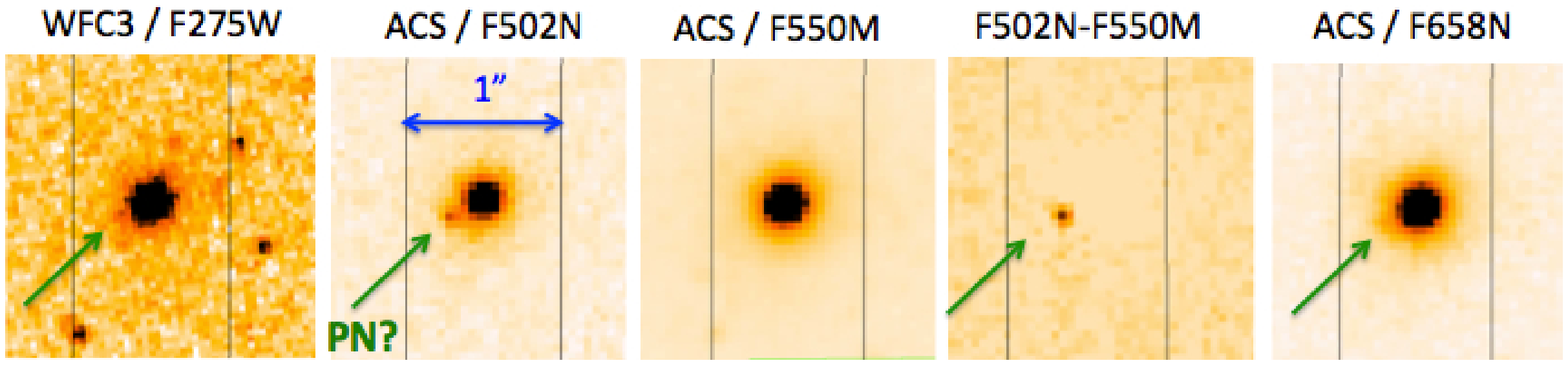}
\caption{HST images of cluster CL~58 in different bands: F275W (NUV) from GO program 13364 (PI Calzetti), and F502N ([O~III]), F550M, 
and F658N (H$\alpha$) from GO program 10522 (PI Calzetti). The continuum-subtracted [OIII] image (F502N$-$F550M) well shows the presence of a PN candidate in cluster CL~58. 
The PN is also detected in the NUV and (with a low signal-to-noise) in H$\alpha$. Superimposed on the cluster is the 1'' - wide slit used in our MODS  observation. \label{cl58_pn}}
\end{figure*}

As discussed in Section~\ref{data_reduction}, the  emission lines observed in the MODS spectrum of cluster CL~58 are not due to the extended ionized gas present in NGC~4449, but instead appear to originate within the cluster itself. In order to investigate the origin of this centrally-concentrated emission, we inspected HST images in different bands, with a spatial resolution $\sim$10 times better than the typical seeing of our observations. In particular, we show in Figure~\ref{cl58_pn} images of cluster CL~58  in F275W (NUV) from GO program 13364 (PI Calzetti), and in F502N ([O~III]), F550M (continuum near [OIII]),  
and in F658N (H$\alpha$) from GO program 10522 (PI Calzetti). These images, and in particular the continuum-subtracted [OIII] image (F502N$-$F550M), reveal the presence of 
a source with strong emission in [O~III] at a projected distance of $\sim$0.25'' from the cluster center; this source is also visible in the NUV image and in H$\alpha$ (with a lower contrast than in [O~III]), while it is not visible in the F550M continuum image. All these properties suggest that it may be a PN belonging to the cluster, although we can not exclude a chance superposition of a ``field'' PN at the cluster position. The [O~III] emission inferred from our MODS spectrum of CL~58 is $\sim3\times10^{-16} erg/s/cm^2$ (see Table~\ref{emission_tab}), compatible with the brightest PNe detected in NGC~4449 \citep{annibali17}.

Planetary nebulae in globular clusters are rare. Only four PNe are known in Galactic GCs so far \citep{pease28,gillett89,jacoby97}. 
\citet{jacoby13} identified 3 PN candidates in a sample of 274 M~31 GCs. \citet{bond15} analyzed HST data of 75 extragalactic GCs in different Local Group galaxies (LMC, M~31, M~33, NGC~147, NGC~6822) to search for PNe, and found only two (doubtful) candidates in the vicinity of the M31 globular cluster B086. A PN was discovered in the NGC~5128 (CenA) globular cluster G169 by  \citet{minniti02}, and one in the Fornax GC H5 by \citet{larsen08}. 
We visually inspected the continuum-subtracted [O~III] image of NGC~4449 to search for additional PN candidates associated with ``old'' ($>$1 Gyr, from integrated colors) clusters in  the \citet{anni11} catalog 
(only 20 of them fall within the F502N field of view), but could not find any. Therefore we are left with one PN detection out of a sample of 20 old clusters in NGC~4449. To our knowledge, CL~58 is the only known dwarf-irregular globular cluster to host a candidate PN. 

According to stellar evolution,  globular clusters as old as the Galactic ones should be unable to host PNe, because in such old populations the masses of stars that are today transiting between the AGB and the white dwarf (WD) phase are smaller than $\sim0.55 M_{\odot}$ \citep[see][and references therein]{jacoby17}. Some sort of binary interaction has  therefore been suggested to explain the 
presence of the (few) PNe detected in old Galactic globular clusters \citep{jacoby97}. On the other hand, globular clusters with intermediate rather than extremely old ages are expected to be richer in PNe than  Milky Way GCs. For the NGC~4449 globular clusters, ages are derived with very large uncertainties; nevertheless, as we will discuss in 
Section~\ref{discussion}, the derived chemical paths tend to exclude a rapid and early cluster formation as in the case of the Milky Way. The idea that NGC~4449 lacks, or has very few, old clusters is reinforced by our detection of one PN out of 20 analysed clusters, a frequency higher than that derived in the MW or in M~31.

\section{Dynamical properties of NGC~4449} \label{dynamics}

We can use the cluster velocities to probe the properties of the cluster population, and of NGC\,4449 as a whole.

\subsection{Systemic velocity and cluster velocity dispersion} \label{meandispersion}

Using the central coordinates and distance of NGC\,4449 given in Section~\ref{intro}, along with the cluster coordinates in Table~\ref{rvel}, we calculate the projected distance of each cluster from the centre of NGC\,4449. The furthest cluster in our sample lies as a projected distance $\Rmax$=2.88~kpc from the centre of NGC\,4449.

We augment our data with the cluster sample from \citet{strader12}, removing any duplicates. This provides an additional 23 clusters. These clusters come partly from the same field as our sample, with some additional clusters detected in SDSS data. There are very few clusters outside the range of our field, and the sample is likely to be incomplete. As such, we elect to include only the 7 additional clusters inside $\Rmax$; this leaves us with a sample of 19 clusters in total. Another reason for this choice is that, in the next section, we will use the cluster velocities to estimate the mass of NGC\,4449; including the clusters outside $\Rmax$ could significantly bias our mass estimates as they would hinge on just one or two distant clusters.

We use a simple maximum-likelihood estimator to evaluate the mean velocity $v = 203 \pm 11$~\kms\ and velocity dispersion $\sigma = 45 \pm 8$~\kms\ of the cluster sample. We will adopt the mean as the systemic velocity of NGC\,4449 and subtract it from the individual cluster velocities in subsequent analysis.
The systemic velocity is in good agreement with the systemic velocity estimate of $205 \pm 1$~\kms\ from \citet{hunter05} and of $204 \pm 2$~\kms\ from \citet{strader12}, though smaller than earlier estimates of $\sim$214~\kms\ from \citet{bajaja94} and \citet{hunter02}.

Velocity measurements of HI gas find that the velocity dispersion of NGC\,4449 varies from 15-35~\kms\ through the galaxy \citep{hunter05}, slightly lower than we find here for the central globular cluster population. H$\alpha$ measurements find even higher dispersions, with a global average dispersion of $31.5 \pm 10$~\kms\ measured by \citet{valdez02}, though they note that the dispersion increases to $\sim 40$~\kms\ in the bar region near the centre, in good agreement with our cluster measurement. These dispersions are generally higher than would be expected, which is indicative of a system that is highly perturbed due to intense star formation \citep{hunter98}.

\subsection{Dynamical mass estimate}
\label{dynamicalmass}

We can also use the radial velocities along with the positions of the clusters to estimate the total mass of NGC\,4449 within the region spanned by the clusters, that is inside $\Rmax$. To do this, we use the tracer mass estimators introduced in \citet{watkins10}, which are simple and yet remarkably effective. There are a number of estimators depending on the type of distances and velocities available; given the data we have, we use the estimator requiring projected distances and line-of-sight velocities here.

The tracer mass estimators assume that, over the region of interest, the tracer sample has a number density distribution that is a power-law with index $-\gamma$, has a potential that is a power-law with index $-\alpha$, and has a constant anisotropy $\beta$. We assume that the sample is isotropic and thus that $\beta = 0$. The other two parameters require further consideration.

To estimate the power-law index $\gamma$ for the number density distribution, we combined the globular cluster samples from \citet{anni11} and \citet{strader12}. As we are only concerned with the number density distribution here, we can use all clusters, regardless of whether or not they also have velocity measurements. With duplicates removed, this leaves a sample of 157 clusters. Again using the central coordinates and distance of NGC\,4449, we calculate the projected distance of each cluster from the centre and then calculate the projected cumulative number profile. To estimate $\gamma$, we fit to this cumulative profile only in the region where we have velocity data ($0.17<R<2.88$~kpc) assuming an underlying power-law density and obtain a best-fitting power-law index $\gamma = 2.18 \pm 0.01$, where the uncertainty here indicates the uncertainty in the fit for the particular model we have assumed and does not account for any additional uncertainties.

To get a better estimate of the uncertainty in $\gamma$, we also fit a double power-law, where we assume that the power-law index is $\gamma_\mathrm{in}$ for $R \le R_\mathrm{break}$ and is $\gamma_\mathrm{out}$ for $R > R_\mathrm{break}$ for some break radius $R_\mathrm{break}$, to the full cluster position sample. The best-fit gives $\gamma_\mathrm{in} = 1.78 \pm 0.01$ and $\gamma_\mathrm{out} = 4.41 \pm 0.03$ with $R_\mathrm{break} = 1.59 \pm 0.01$~kpc. So we will use $\gamma = 2.18$ to obtain our best mass estimate, but use the range $1.78 \le \gamma \le 4.41$ to provide uncertainties on the mass estimate.

\begin{figure}
    \centering
    \includegraphics[width=0.9\linewidth]{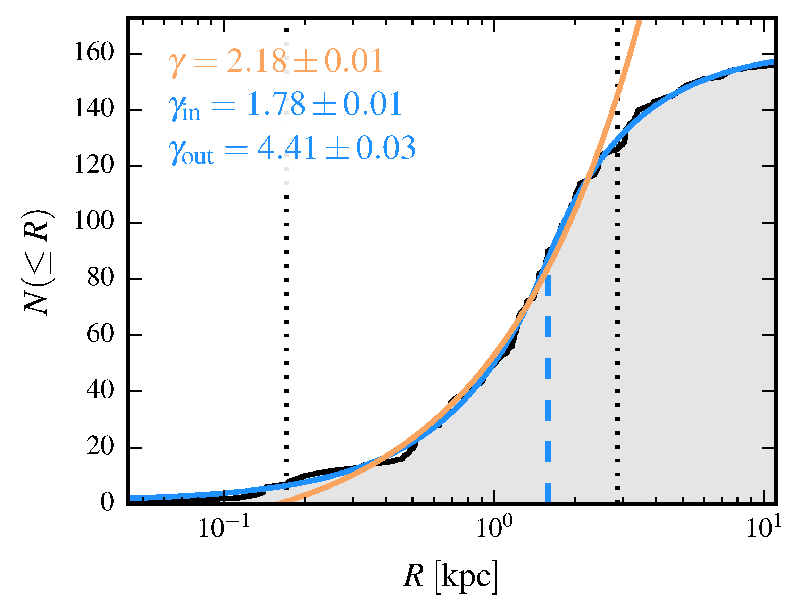}
    \caption{The cumulative number density profile of globular clusters in NGC\,4449. The black line shows the profile measured from the combined catalogues of \citet{anni11} and \citet{strader12}. The orange line shows the best-fitting profile to the region of interest (marked by vertical dotted lines) assuming that the underlying density profile is a power-law. The index of the power-law $\gamma$ is given in orange the top-left corner. The blue line shows the best-fitting profile to the whole sample assuming that the underlying density profile is a double power-law with an inner index of $\gamma_\mathrm{in}$ and an outer index $\gamma_\mathrm{out}$ (shown in blue in the top-left corner). The radius at which the power-law index changes from inner to outer is marked by the blue vertical dashed line. We use the single fit as the best estimate of the power-law in this region and use the indices from the double fit to provide uncertainties.}
    \label{cumulative_number}
\end{figure}

Figure~\ref{cumulative_number} shows the results of these fits. The black line shows the cumulative number density profile of the data, the orange line shows the best-fitting single power-law in the region of interest (the boundaries of which are marked by the vertical dotted lines), the solid blue line shows the best-fitting double power-law to the whole sample, and the dashed blue line marks the break radius for the double power-law fit. The best-fitting power-law indices for the fits are also given.

To estimate the power-law index $\alpha$ of the potential, we assume that NGC\,4449 has a baryonic component embedded in a large dark matter (DM) halo. The power-law slope of the potential will depend on which of these components dominates inside $\Rmax$, or indeed, if both make significant contributions to the potential. If the baryonic component is dominant so that mass follows light, then $\alpha = \gamma - 2$ for $\gamma \le 3$ and $\alpha = 1$ for $\gamma > 3$. For the range of $\gamma$ we found above, this would imply a range of $-0.22 \le \alpha \le 1$ with a best estimate of $\alpha = 0.18$.

To estimate $\alpha$ for the case where the DM component is dominant, we assume that the DM halo of NGC\,4449 is \citet{nfw} (hereafter NFW), but with unknown virial radius and concentration. However, although we do not know the exact halo parameters, we can still place some constraints on these values. The virial radius is likely to be smaller than the Milky Way's value of $r_\mathrm{vir} \sim 258$~kpc, and the concentration is likely to be between the Milky Way's value of $c \sim 12$ and the typical concentration of dwarf spheroidals $c \sim 20$. Guided by these constraints, we try a grid of virial radii ($80 \le r_\mathrm{vir} \le 200$~kpc) and concentrations ($10 \le c \le 20$) and find the power-law that best fits the slope of the potential in the region of interest for each halo.

\begin{figure}
    \centering
    \includegraphics[width=0.9\linewidth]{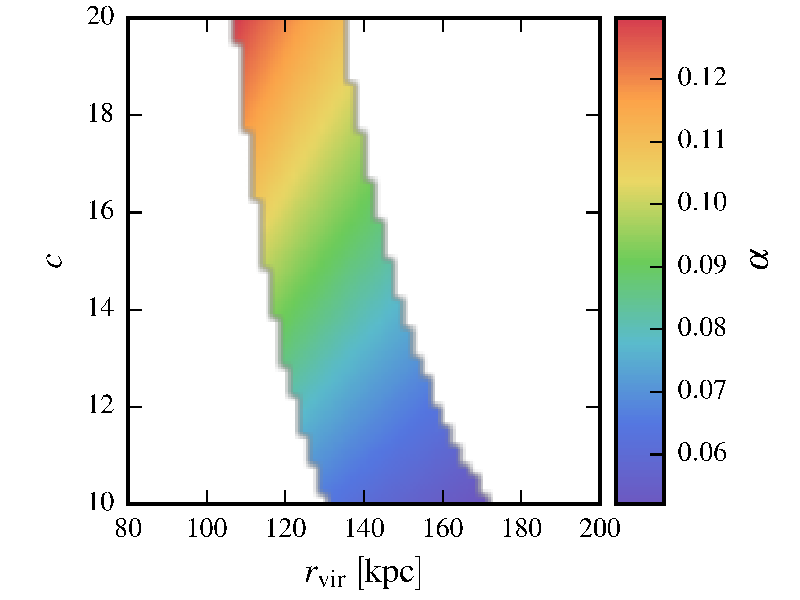}
    \caption{The variation in potential slope $\alpha$ for a grid of DM halos assumed to be NFW in shape and defined by a concentration $c$ and a virial radius $r_\mathrm{vir}$. To each halo, we fit a power-law over the range of interest for our mass estimation -- the colour bar shows how the index $\alpha$ of the best-fitting power-law changes from halo-to-halo. We only show halos with circular velocities at 15.7~kpc of $84 \pm 10$~\kms\ to force consistency with HI observations. The white regions are halos with circular velocities outside of this range.}
    \label{halogrid_alpha}
\end{figure}

To further narrow down the choice of likely halos, we turn to HI measurements, which find a rotation speed, corrected for inclination, of 84~km\,s$^{-1}$ at 15.7~kpc \citep{bajaja94,martin98}. If we insist that the circular velocity at 15.7~kpc must be within 10~km\,s$^{-1}$ of the H$\alpha$ measurements, then this restricts the virial radius and concentration combinations allowed. Figure~\ref{halogrid_alpha} shows how $\alpha$ varies over the halo grid for the ``allowed'' halos only. These remaining halos follow a curve in the $r_\mathrm{virial} - c$ plane, and predict $0.05 \lesssim \alpha \lesssim 0.13$ with a mean $\alpha \sim 0.09$, which, we note, is much narrower than the mass-follows-light case.

However, we must still consider whether the DM is the dominant mass component inside $\Rmax$ or if the baryonic mass dominates the form of the potential. Using $\gamma = 2.18$ as previously determined, the total mass enclosed within $\Rmax$ for this range of $\alpha$ values shows little variation (only $\sim$1\%). However, the total DM mass inside $\Rmax$ changes significantly from halo-to-halo, such that the DM is approximately half of the total mass at its lowest and apparently exceeds the total mass at its highest (we choose not to rule out these halos as a larger $\gamma$ would alleviate this problem). Overall, we conclude that the DM makes a significant or dominant contribution to the mass within $\Rmax$ and so it would be incorrect to assume that mass-follows-light. So we will use the mean value of $\alpha = 0.09$ found from the NFW fitting to obtain our best mass estimate, but, to be conservative, use the broader range $-0.22 \le \gamma \le 1$ to provide uncertainties on the mass estimate.

\begin{figure}
    \centering
    \includegraphics[width=0.9\linewidth]{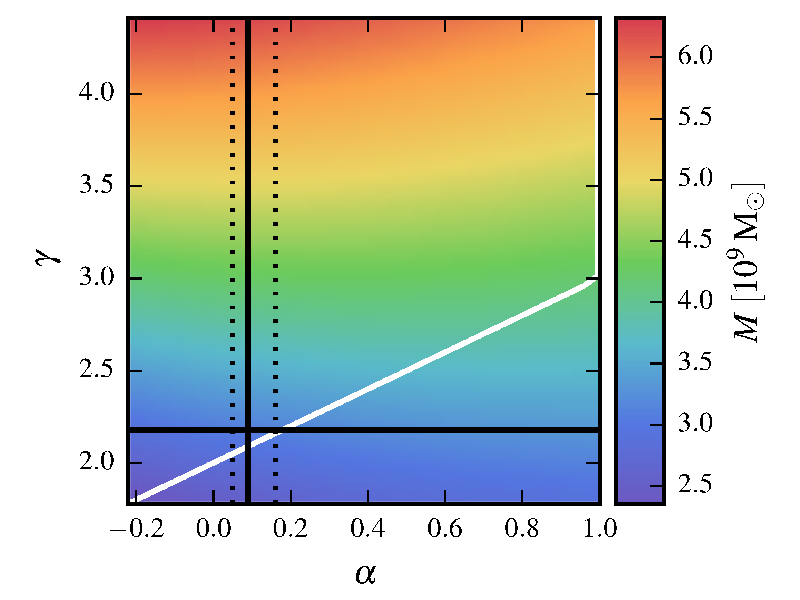}
    \caption{The variation in estimated mass inside 2.88~kpc as a function of potential slope $\alpha$ and density slope $\gamma$. The range of $\gamma$ was set by fitting to existing globular cluster data. The range of $\alpha$ was set by assuming that mass follows light, in which case $\alpha = \mathrm{min}(1, \gamma - 2)$ (marked by the white line), as this range was larger than that estimated from considering DM halo models (highlighted by vertical dotted lines). The solid black lines show our best estimates for $\alpha$ and $\gamma$, and their intersection shows the adopted best mass estimate. The minimum and maximum masses in this grid set our uncertainties.}
    \label{massgrid}
\end{figure}

Now that we have estimates for $\alpha$ and $\gamma$, we can proceed with the mass estimates. As discussed, we will use our best estimates of $\alpha = 0.09$ and $\gamma = 2.18$, to get a ``best'' mass estimate; and to estimate the uncertainty in the mass, we will calculate the mass estimate across $-0.22 \le \alpha \le 1$ and $1.78 \le \gamma \le 4.41$. Figure~\ref{massgrid} shows the results of these calculations. For each $\alpha$-$\gamma$ combination across the grid, we estimate the mass inside $\Rmax$. The solid black lines shows our adopted ``best'' $\alpha$ and $\gamma$ values. The dotted black lines show the range of $\alpha$ predicted from the NFW halo fitting. The white line traces the results if we were to assume that mass-follows-light. We see that the mass estimate is dominated by the assumed value of $\gamma$ and is much less sensitive to the value of $\alpha$. Overall, we estimate the mass of NGC\,4449 inside 2.88~kpc to be $M(<2.88\,\mathrm{kpc}) = 3.15^{+3.16}_{-0.75} \times 10^9$~M$_\odot$. The implied circular velocity at $\Rmax = 2.88$~kpc is $69^{+29}_{-9}$~\kms. By comparison, \citet{martin98} estimated a rotation of $\sim$40~\kms\ at $\sim$3~kpc using HI observations, which is somewhat lower than our estimate here. However, the significant velocity dispersion of the HI (see Section~\ref{meandispersion}) implies that the HI rotation speed should indeed be below the circular velocity to maintain hydrostatic equilibrium.

\citet{hunter02} found a minimum in the HI distribution at the centre of NGC\,4449, and estimated that the HI mass inside 2~arcmin ($\sim$2.2~kpc) is just $3 \times 10^8$~M$_\odot$. Assuming that the mass increases linearly with radius, the HI mass inside $\Rmax$ would be $\sim 4 \times 10^8$~M$_\odot$, so the HI makes up a little more than 10\% of the total mass near the centre. The total mass was estimated from HI observations to be $\sim 7.8 \times 10^{10}$~M$_\odot$ inside 30$^{\prime}$ ($\sim$ 33~kpc). If we assume that NGC\,4449 has a flat rotation curve, and thus that the mass increases linearly with radius, our mass estimate would imply a total mass $\sim 3.6^{+3.6}_{-0.9} \times 10^{10}$~M$_\odot$ within 30$^{\prime}$, which is inconsistent with the HI measurements, even given our rather generous upper mass limit. Furthermore, our estimate is likely to be an overestimate as the assumption of a flat rotation curve is likely reasonable near the centre but may break down towards larger radii, thus creating more tension with the mass implied by HI data.

NGC\,4449 is often considered an analog of the Large Magellanic Cloud (LMC). By comparison, \citet{marel14} find a mass $M(<8.7\,\mathrm{kpc}) = 17 \pm 7 \times 10^{9}$~M$_\odot$ for the LMC. Again assuming that the mass increases linearly with radius, then our mass measurement would imply a mass of $\sim 9.5 \times 10^{9}$~M$_\odot$ at 8.7~kpc, with the uncertainties encompassing $7-19 \times 10^{9}$~M$_\odot$. This estimate is lower than the LMC estimate, but still consistent at the upper end of the uncertainty range.

\section{Discussion} \label{discussion}

From our stellar population study of old clusters in NGC~4449, we find that the large majority of them have solar or sub-solar [$\alpha$/Fe] and [N/$\alpha$]  ratios, making them clearly different with respect to Galactic GCs.
These chemical properties have strong implications on our understanding of the formation of globular clusters in this irregular galaxy, which we discuss in the following.


\subsection{[$\alpha$/Fe] ratios}

It is well known that the [$\alpha$/Fe] ratio quantifies the relative importance of high mass stars versus intermediate/low mass stars 
to the enrichment of the interstellar medium (ISM)
and it is therefore tightly linked to both the stellar initial mass function (IMF) and to the SF timescale  \citep[e.g.][]{greggio83,tornambe86}.
In fact, $\alpha$-elements (O, Ne, Mg, Si, S, Ar, Ca, Ti) are mainly  synthesized by massive stars  and restored into the interstellar medium (ISM) on short SF timescales ($\lesssim$50 Myr), while Fe, mostly provided by SNIa, is injected into the ISM on longer timescales. Soon as SNIa start to contribute, they dominate the iron enrichment and [$\alpha$/Fe] inevitably decreases: \citet{mr01} estimated that the time in maximum enrichment by SNIa varies from $\sim$40-50 Myr for an instantaneous starburst 
to $\sim$0.3 Gyr for a typical elliptical galaxy to $\sim$4-5 Gyr for a spiral galaxy.  Therefore, under the assumption of a universal IMF,  super-solar  [$\alpha$/Fe] ratios can be interpreted as the signature of an early and rapid star formation (e.g., in giant elliptical galaxies, or in the halo's field stars and globular clusters of spiral galaxies), while sub-solar [$\alpha$/Fe] indicate a prolonged star formation, more typical of dwarf systems \citep[e.g.][]{mt85,lanfranchi03}. In dwarf galaxies, galactic winds powered by massive stars and SN~II explosions may also play an important role in lowering the [$\alpha$/Fe] value through expulsion of the newly produced $\alpha$-elements: \citet{marconi94} showed that sub-solar [$\alpha$/Fe] ratios can be obtained  in dwarf galaxies by assuming differential galactic winds that are more efficient in removing the $\alpha$-elements than Fe since activated only during the bursts where most of the $\alpha$-elements are formed. Alternatively, different [$\alpha$/Fe] ratios could be due to a non-universal IMF: for instance, it has been suggested that the IMF could have been skewed towards massive stars in elliptical galaxies and in the Bulge, producing the observed super-solar $\alpha$/Fe ratios,  and towards low-mass stars in dwarf galaxies \citep[e.g.][]{weidner13,yan17}.

From an observational point of view, chemical abundances of individual red giant branch stars have been extensively derived in the nearest dwarf spheroidal galaxies 
\citep[dSphs; e.g.,][]{shetrone98,shetrone01,tolstoy03,kirby11,lemasle14,norris17,apogee17}.  These studies have shown that each dSph starts, 
at low [Fe/H], with super-solar [$\alpha$/Fe] ratios, similar to those in the Milky Way halo at low metallicities, and then the [$\alpha$/Fe] ratio evolves down to lower values than are seen in the Milky Way at high metallicities \cite[see][for a review]{tht09}. The ``knee'', i.e.  the position where [$\alpha$/Fe] starts to decrease, depends 
on the particular SFH of the galaxy and occurs at larger [Fe/H] for a more rapid star formation and a more efficient chemical enrichment. The general finding in dSphs that relatively metal rich ([Fe/H]$>-1$) stars are deficient in $\alpha$-elements compared to iron suggests that the most recent generations of stars were formed from an ISM relatively poor in the ejecta from SN~II. 

The observational picture is far less complete for dwarf irregular galaxies because of their larger distance from the Milky Way compared to dSphs (with the obvious 
exceptions of the LMC and the SMC). Chemical abundance  determinations in  these systems are mainly limited to H~II regions \citep[e.g.,][]{it99,kniazev05,magrini05,vanzee06,guseva11,haurberg13,berg12} and to a few supergiant stars \citep{venn01,venn03,kaufer04,leaman09,leaman13}, which probe a look-back time of at most $\sim$10 Myr. Chemical abundance studies of planetary nebulae  \citep[e.g.,][]{magrini05,pena07,rojas16,flores17,annibali17} and unresolved star clusters 
\citep[e.g.,][]{strader03,puzia08,sharina10,hwang14} provide a valuable tool to explore more ancient epochs and have been attempted in a few dwarf irregulars. 
The H~II region and young supergiant tracers indicate typically low [$\alpha$/Fe] ratios, as expected in galaxies that have formed stars at a low rate over a long period of time and  where galactic winds may have possibly contributed to the expulsion of the $\alpha$-elements. Interestingly, also integrated-light studies of globular clusters in dwarf irregulars provide  solar or slightly sub-solar [$\alpha$/Fe] values, in agreement with our results. 
In NGC~4449 we find only a minority of clusters (3 in our sample) with super-solar [$\alpha$/Fe] ratios at [Fe/H]$\lesssim-1.2$, while the majority of clusters (8 in our sample) 
display intermediate metallicities ($-1\lesssim$[Fe/H]$\lesssim-0.5$) and sub-solar [$\alpha$/Fe] values. This indicates that the NGC~4449's clusters  
have  typically formed from a medium already enriched in the products of SNe type Ia and relatively poor, compared to the solar neighborhood, in the products of massive stars and SN~II (either because of the slower star formation history and/or because of preferential loss of $\alpha$-elements due to galactic winds).
From our data there are hints that the {\it knee}  occurs at [Fe/H]$\le-1.2 - -1.0$, although a larger cluster sample  would be needed to reinforce this result. In particular, we do not observe a very metal  poor (i.e, $[Fe/H]<-1.5$), $\alpha$-enhanced cluster population in our sample.
Whether such a population is present at larger galactocentric distances than those sampled by our data   \citep[e.g.,][identified possible cluster candidates in the outer halo of NGC~4449]{strader12}, or if such a component is absent due to the lack of a major star formation event at early times in NGC~4449, can not be established from our data. 
Indeed, resolved-star color-magnitude diagrams indicate a low activity at ancient epochs in dIrrs as opposed to an early peak in the star formation history of dSphs \citep[e.g.][]{monelli10,weisz14,skillman17}; this behaviour could naturally explain the different chemical paths observed in dSphs and  dIrrs. The idea that the typical globular cluster population in NGC~4449 is younger than in the MW is reinforced by our serendipitous detection of a PN in cluster CL~58 (see Section~\ref{cl58_section}). 

A  useful comparison is that between NGC~4449 and the LMC, the irregular whose star cluster system is the best studied. 
The LMC globular clusters span a wide age/metallicity range, with both old, metal-poor and young, metal-rich objects, due to its complex star formation history. Chemical abundances of individual RGB stars in LMC clusters have been derived by several authors \citep[e.g.][]{hill00,j06,mucciarelli10}. These studies show that old LMC clusters display a behavior of [$\alpha$/Fe] as a function of [Fe/H] similar to the one observed in the Milky Way stars, with old LMC clusters having  [$\alpha$/Fe]$\sim0.3$; these clusters should therefore have formed during a rapid star formation event that occurred at early times, when the ISM was not yet significantly enriched by the products of SNIa. However, the majority of clusters in the LMC exhibit relatively ``young'' ages and intermediate metallicities  1-3 Gyr, [Fe/H]$\sim$-0.5 coupled with low [$\alpha$/Fe] values, more chemically similar to the clusters in NGC~4449.  A possible explanation is that the LMC has formed the majority of its GCs from a medium already 
enriched by SNIa. 

Two of the most luminous clusters in our sample, namely clusters CL~77 and CL~79, were also studies by \citet{strader12} through spectroscopy (their clusters B15 and B13, respectively). 
These clusters appear peculiar in that they are quite massive ($M_{\star}\sim2\times10^6 M_{\odot}$,  comparable to the mass of $\omega$Cen in the MW, that is thought to be the remnant nucleus of an accreted dwarf galaxy) and  elongated ($1-b/a\sim0.2-0.3$); cluster CL~77 is furthermore associated in projection with two tails of blue stars whose shape is reminiscent of tidal tails, and has therefore been suggested to be the nucleus of a former gas-rich satellite galaxy undergoing tidal disruption by NGC 4449 \citep{annibali12}.  
For clusters  CL~77 and CL~79, \citet{strader12} derived ages of  11.6$\pm$1.8 Gyr and  7.1$\pm$0,5, a common metallicity of [Fe/H] $=-1.12\pm0.06$ dex, and [$\alpha$/Fe] values of $-0.2\pm0.1$ and $-0.1\pm0.1$, respectively; these results are marginally consistent with our age estimates of 12$\pm$2 Gyr and 11$\pm$2 Gyr,  [Fe/H] metallicities of $-0.8\pm0.2$ dex and  $-0.9\pm0.2$ dex, and [$\alpha$/Fe] ratios of $-0.5\pm0.2$ and $-0.4\pm0.2$, respectively. While the relatively poor signal-to-noise of the \citet{strader12} spectra did not allow them to derive the stellar population parameters for other clusters in NGC~4449 than CL~77 and CL~79, our study shows that the presence of sub-solar [$\alpha$/Fe] values is a typical characteristic of the NGC~4449's  clusters and not just a peculiarity of clusters  CL~77 and CL~79. 

\subsection{[N/$\alpha$] ratios}

Besides displaying sub-solar [$\alpha$/Fe], the  majority of clusters in NGC~4449 appear peculiar also in their CN content, as quantified by the CN$_1$ and CN$_2$ indices: compared to 
Galactic globular clusters (or to M31 globular clusters), they show lower CN indices at the same metallicities. The CN absorption strength is sensitive to the abundances of both carbon and nitrogen; however, the CN lines observed in  Galactic globular clusters are well reproduced by models in which N, rather than C, is enhanced with respect to the solar composition \citep[e.g.][]{tmb03,puzia05}: the reason is that a C enhancement would result also into a significant increment of the C$_2$4668 and Mg$_1$ indices, in disagreement with the observations. Using Lick indices, \citet{tmb03} derived [N/$\alpha$]$\sim$0.5 for Galactic globular clusters. 
   
Recent progress in studies of globular clusters has shown that they contain multiple stellar populations \citep[e.g.,][]{piotto15}: at  least two populations of stars are present, one with the same chemical pattern as halo-field stars, and  second ones which are enhanced in helium, nitrogen and sodium and depleted in carbon and  oxygen 
\citep[e.g.,][]{gratton12}.  A possible scenario to explain this multiplicity is that the second populations have formed from a material enriched from the ejecta of intermediate-mass asymptotic-giant branch stars  \citep[e.g.][]{renzini15,dantona16}. 
While the presence of multiple populations seems so far to be ubiquitous in globular clusters, the fraction of  first over second-generation stars varies from cluster to cluster, with the first generation of stars typically being the minority;  therefore, we would expect the second, N-enriched generations of stars to dominate the integrated cluster light and to drive the large N-enrichment observed in the integrated spectra of Galactic  globular clusters. The scenarios to interpret the cluster multiple populations are still very uncertain though, and none of them provide a good explanation for all their observed properties yet \citep{bastian17}.

 With the exception of  CL~72, CL~75, and CL~58, the CN$_1$ and CN$_2$ indices for  the clusters in NGC~4449 lie slightly below the models with [N/$\alpha$], [C/$\alpha$]$=$0. Unfortunately, our data do not allow us 
to establish if the low CN absorption is due to either a N or to a C depletion: in fact, as already discussed in Section~\ref{nitrogen_carbon}, 
the C$_2$4668 and Mg$_1$ indices, which are highly sensitive to C and could in principle allow for a discrimination between the two effects, show a very large scatter in our sample and are not well calibrated. Under the assumption that the low observed CN strength is due to a N depletion,  we computed N/$\alpha$ ratios for the clusters and found 
$-0.5\lesssim[N/\alpha]\lesssim-0.1$ (while clusters CL~72, CL~75, and CL~58 show highly super-solar  ratios of [N/$\alpha]>0.5$).
We stress that, although our data do not permit discrimination between N or C depletion, we can definitively exclude for the majority of  clusters in NGC~4449 the presence of a major N-enriched component similar to that observed for old Galactic globular clusters.  However, we can not exclude the presence of intrinsic light-element variations, although to a different extent with respect to MW GCs \citep[see e.g. the case of cluster NGC~419 in the SMC,][]{martocchia17}.

\subsection{Comparison between gas and star metallicities}

From our spectroscopic study, we derived cluster total metallicities in the range $-1.3 \lesssim \log Z/Z_{\odot} \lesssim -1.0$ (see Table~\ref{agezafe_tab}), for an adopted solar metallicity of $Z_{\odot}=0.018$. Since the total metal fraction is mainly driven by oxygen, it is sensible to compare these values with oxygen abundance determinations in NGC~4449' s H~II regions, which trace the present-day ISM composition. \citet{berg12} and \citet{annibali17} derived metallicities in the range $8.26\lesssim 12 + \log(O/H) \lesssim8.37$; this translates, for an assumed solar oxygen abundance of  
$12+\log(O/H)_{\odot}=8.83\pm0.06$ \citep{grevesse98}\footnote{We adopt $12+\log(O/H)_{\odot}=8.83\pm0.06$  and  $Z_{\odot}=0.018$ instead of the lower, more recent estimates of $12+\log(O/H)_{\odot}= 8.76\pm0.07$ and $Z_{\odot}=0.0156$ from \citet{caffau08,caffau09} to be consistent with the solar abundance values adopted in our SSP models  (see Section~\ref{stpop}).},  into $-0.57\lesssim\log((O/H)/(O/H)_{\odot})\lesssim-0.46$. This result indicates a $\gsim$0.5 dex metal enrichment in NGC~4449 within the last $\sim$10 Gyr, which is the typical age of the clusters in our sample. A spectroscopic estimate of NGC~4449' s stellar metallicity was also derived by  \citet{kar13}: they inferred an average  $\sim$1/5 solar metallicity for the stellar population older than 1 Gyr, which is ``qualitatively'' consistent with the $\sim$1/10 solar metallicity derived in this paper for our $\sim$10 Gyr old stellar clusters. In a future paper (Romano et al. in preparation) we will use all the available information on the chemical properties of the stellar and gaseous components in NGC~4449 to run chemical evolution models that will allow us to produce a self-consistent scenario for the evolution of this galaxy.

\section{Conclusions} \label{conclusions}

We acquired intermediate-resolution (R$\sim$1000) spectra in the range $\sim$ 3500$-$10,000 \AA \ with the MODS instrument on the LBT  for a sample of 14 star clusters in the  irregular galaxy NGC~4449. The clusters were selected from the sample of 81 young and old clusters of \citet{anni11}. With the purpose of studying the integrated-light  stellar population properties, we derived for our clusters Lick indices in the optical and the CaII triplet index in the near-infrared.  The indices were then compared with simple stellar population models to derive ages, metallicities, [$\alpha$/Fe] and [N$/\alpha$] ratios. Of the 14 clusters observed with MODS, 3 are affected by a major contamination  from the diffuse ionized gas in NGC~4449 (in particular, one of these is a $\sim$few hundred Myr old cluster). Therefore, we are left with a sub-sample of 11 clusters for which we could 
perform a reliable stellar population analysis. For this sub-sample, the main results are:

\begin{itemize}
\item The clusters have intermediate metallicities, in the range  $-1.2\lesssim$[Fe/H]$\lesssim-0.7$; the ages are typically older than $\sim$9 Gyr, although determined with large uncertainties. No cluster with  iron metallicity as low ($-2\lesssim$[Fe/H]$\lesssim-1.2$)  as in Milky Way globular clusters is found in our sample.  

\item The  majority of clusters exhibit sub-solar $\alpha$/Fe ratios (with a peak at $[\alpha/Fe]\sim-0.4$), suggesting 
that they formed from a medium already enriched in the products of Type Ia supernovae.  Sub-solar [$\alpha$/Fe] values are expected in galaxies that formed stars 
inefficiently and at a low rate and/or  where galactic winds possibly contributed to the expulsion of the $\alpha$-elements.

\item Besides the low [$\alpha$/Fe], {the majority of clusters in NGC~4449 appear} also to be under-abundant in CN compared to Milky Way halo globular clusters. 
A possible explanation is the lack of a major contribution from N-enriched, 
second-generation stars as those detected in the old, metal-poor galactic globular clusters.  Intrinsic light-element variations may still be present within NGC~4449' s GCs, but to 
a different extent with respect to MW clusters.

\item We report the serendipitous detection of a PN within cluster CL~58 out of a sample of 20 ``old''  ($>$1 Gyr, from integrated colors) clusters in NGC~4449 covered by ACS/F502N and ACS/F555M images. PNe in old MW and M31 globular clusters are extremely rare, and our result reinforces the idea that the cluster population in NGC~4449 is typically younger than in these two giant spirals.

\item We use the cluster velocities to infer the dynamical mass of NGC~4449. We estimate the mass of NGC\,4449 inside 2.88~kpc to be 
M($<$2.88 kpc)=$3.15^{+3.16}_{-0.75} \times 10^9~M_\odot$;  the upper mass limit  within 30$^{\prime}$ is  $\sim 3.6^{+3.6}_{-0.9} \times 10^{10}$~M$_\odot$, significantly lower than the mass derived in the literature  from HI data.

\end{itemize}

\section*{Acknowledgements}

This work was based on LBT/MODS data. 
The LBT is an international collaboration among institutions in the United States, Italy and Germany. LBT Corporation partners are: the University of Arizona on behalf of the Arizona Board of Regents; Istituto Nazionale di Astrofisica, Italy; LBT Beteiligungsgesellschaft, Germany, representing the Max-Planck Society, the Leibniz Institute for Astrophysics Potsdam, and Heidelberg University; the Ohio State University, and the Research Corporation, on behalf of the University of Notre Dame, University of Minnesota and University of Virginia.
We acknowledge the support from the LBT-Italian Coordination Facility for the execution of observations, data distribution and reduction.
F. A. and M. T acknowledge funding from INAF PRIN-SKA-2017 program 1.05.01.88.04. We thank the anonymous referee for his/her very nice report. 

\appendix{}

\section{Lick and Ca~II triplet indices derived for globular clusters in NGC~4449}

\begin{landscape}
 \begin{table}
  \caption{Lick and Ca~II triplet indices derived for globular clusters in NGC~4449.}
  \label{indices_table}
  \begin{tabular}{lcccccccccccccccccc}
\hline
Name &   CN$_1$ & CN$_2$    & Ca4227 & G4300 &  Fe4383 & Ca4455 &   Fe4531 &  C$_2$4668 &    H$\beta$ &  Fe5015  &   Mg1   &    Mg2 &  Mgb &  Fe5270 &  Fe5335 &  Hga &  Hgf &  CaT*     \\
 &   eCN$_1$ & eCN$_2$    & eCa4227 & eG4300 &  eFe4383 & eCa4455 &   eFe4531 &  eC$_2$4668 &    eH$\beta$   & eFe5015  &     eMg1 &   eMg2 &  eMgb &  eFe5270 &  eFe5335 &  eHga &  eHgf &  eCaT*    \\
 \hline         
CL~20     &     -0.10  &      -0.06    &       0.3   &        1.4  &        -0.5   &        0.4     &     0.0    &    -0.9  &       2.0    &     1.7     &    0.042  &       0.078   &      1.1   &      1.3   &      1.1   &           -0.2   &  0.6    &     7.0      \\
CL~20      &     0.01  &        0.02   &       0.2    &       0.5   &        0.7    &       0.3      &     0.5    &     0.9  &        0.3      &     0.7     &    0.009   &      0.009   &      0.3   &      0.4   &      0.4   &            0.5    & 0.3     &    0.8      \\
CL~67      &   -0.13    &      -0.07     &      0.0     &     -2.8      &    -1.1    &      -0.2  &        0.1    &      -6    &       6.4    &    0.0      &   0.04      &   0.08     &    0.7     &    0.4     &    0.1       &       10.1    &  7.3   &     7.0      \\
CL~67      &    0.02    &       0.03    &       0.4      &     0.9       &    1.3     &      0.6    &       1.0     &      2     &      0.6      &    2.0      &   0.02      &   0.02     &    0.8     &    0.9      &   1.0       &        0.7    &  0.4    &     3.0      \\
CL~76      &        -0.081  &       -0.07  &          0.34  &        3.1   &      0.2   &      1.0      &      2.6   &      -0.05   &       2.0    &  2.1      &   0.058    &     0.125    &     1.4      &   1.7    &     1.5    &          -1.6   &  0.1    &     6.5      \\    
CL~76      &         0.005   &        0.01  &         0.08   &        0.2   &      0.2   &       0.2     &      0.1   &       0.34    &       0.1     & 0.3       &  0.006    &     0.004     &    0.1       &  0.1     &    0.1     &          0.2    &  0.1      &   0.3    \\ 
CL~72      &       0.02    &       0.07    &       0.8   &        3.4      &     1.4     &      0.0     &      2.0   &       -0.2   &        1.6  &     3.9       &  0.054     &    0.111    &     1.6    &     1.9     &    1.3     &         -1.7  &   0.4   &      6.9       \\
CL~72      &       0.01    &       0.02     &      0.2    &       0.4     &      0.5     &      0.3      &     0.4    &       0.6  &         0.2      &     0.6       &  0.008     &    0.007    &     0.2    &     0.3     &    0.3     &          0.4   &  0.2   &      0.6       \\
CL~75     &    -0.02      &    0.02       &    0.6    &       3.2    &       1.0     &      0.1    &      1.4    &       1.5      &     1.9      &     2.1       &  0.044     &    0.078    &     0.8     &    1.5    &     1.0       &        0.2   &  0.4    &     3.5      \\
CL~75     &    0.01      &     0.02        &   0.2    &       0.4      &     0.6      &     0.3      &     0.5    &       0.8     &      0.3      &     0.7       &  0.009     &    0.009    &     0.3     &    0.3    &     0.4       &        0.5    & 0.3    &     0.8     \\
CL~79     &         -0.076   &       -0.05      &     0.55    &       2.8     &      1.2     &   0.7   &      2.2     &    1.9     &     2.0   &     2.8       &  0.058     &    0.107    &     1.2     &    1.5    &     1.1       &       -1.2    &  0.8    &     6.4    \\    
CL~79     &          0.005  &         0.01     &      0.08     &      0.2    &       0.2      &   0.2     &    0.2  &       0.3   &      0.1     &     0.4        & 0.006    &     0.004    &     0.1     &    0.1    &     0.1       &        0.2     & 0.1     &    0.3    \\ 
CL~58     &     -0.01  &       0.02    &     0.6    &     2.7     &    0.7    &     1.0     &    2.1     &    0.7   &      2.6   &    1.8      &   0.08     &    0.14     &    1.7     &    1.4      &   1.2       &       -0.6     & 0.8    &     7         \\
CL~58     &      0.02    &     0.02     &    0.3     &    0.6      &   0.9     &    0.4      &   0.7      &   1.0    &     0.5       &    1.0       &  0.01     &    0.01     &    0.5     &    0.5      &   0.6       &        0.7     & 0.4     &    1     \\
CL~52      &       -0.072    &    -0.04     &    0.5  &       2.9   &     0.9   &      1.1  &       1.9   &      0.5    &     2.2    &    3.4       &  0.066    &     0.118    &     1.2    &     1.5     &    1.2      &        -1.2    & 0.9    &     6.9      \\  
CL~52      &        0.009     &    0.01      &   0.2   &      0.4   &     0.4    &     0.2    &     0.3     &    0.6      &   0.2     &    0.5       &  0.008    &     0.007    &     0.2   &      0.3     &    0.3      &        0.3     & 0.2    &     0.6      \\
CL~3       &       -0.077      &  -0.03   &      0.61  &       2.5  &       2.1  &       0.2  &       0.5    &    -0.7    &     1.9      &    3.2       &  0.070     &    0.112     &    1.9      &   1.2     &    1.4       &       -1.1    &  1.2    &     6.2    \\ 
CL~3       &        0.005      &   0.01    &     0.08   &      0.2    &     0.2    &     0.2   &      0.2    &    0.4     &    0.1    &    0.4       &  0.006     &    0.005     &    0.1      &   0.2     &    0.2       &       0.2     & 0.1     &    0.4      \\
CL~77     &       -0.098     &   -0.08     &    0.6   &      3.1   &      1.6    &     1.0  &       2.2      &   3.3   &     2.0      &    1.6       &  0.052    &    0.100    &     1.1     &    1.5     &    0.9      &       -2.4     & 0.7   &      6.1      \\
CL~77     &        0.007      &   0.01      &   0.1    &     0.3    &     0.3  &      0.2   &      0.2       &  0.4     &    0.1      &    0.4       & 0.007     &    0.005    &     0.2     &    0.2     &    0.2      &        0.3     & 0.1    &     0.4     \\
CL~27      &          -0.092    &    -0.07      &   0.7  &     2.0    &     2.2    &     0.8    &     2.0   &      1.8   &      1.9     &    2.9        & 0.069     &    0.117      &   1.5     &    1.7     &    1.3      &        -1.7    & 0.9    &     6.1      \\   
CL~27      &           0.008     &    0.01     &   0.1   &      0.3     &    0.4     &    0.2     &    0.3    &     0.5  &      0.2     &    0.5        & 0.007     &    0.007      &   0.2    &     0.2     &    0.2      &         0.3    & 0.2    &     0.5      \\  
CL~24      &        -0.11    &    -0.09    &    0.0   &      3.1   &      0.6    &    -0.2   &      2.4  &       4.7   &      1.3      &     3         & 0.09    &     0.13    &     1.5     &    0.8      &   0.8       &       -2.6     & 0.1  &       5       \\
CL~24      &         0.02      &   0.02      &   0.3     &    0.6     &    0.8      &   0.4     &    0.7     &    1.0     &    0.4     &     1         & 0.01    &     0.01    &     0.5     &    0.5      &   0.6       &        0.7     & 0.4   &      1      \\
\hline
  \end{tabular}
 \end{table}
\end{landscape}

\end{document}